\documentclass[english,journal=jctcce,manuscript=article,email=true,etalmode=truncate,maxauthors=0]{achemso}

  \usepackage{geometry} 
  \usepackage{stackengine}
  \usepackage{titlesec}

  \usepackage[T1]{fontenc} 
  \usepackage{amsmath}     
  \usepackage{amssymb}
  \usepackage{bm}
  \usepackage{mathrsfs}
  \usepackage{etoolbox}
  \usepackage{environ}
  \usepackage[version=4]{mhchem}
  \usepackage{booktabs}
  \usepackage{adjustbox}
  \usepackage{hhline}

  \usepackage{setspace}    
  \usepackage{float}       
  \newfloat{chart}{htbp}{loc}
  \newfloat{graph}{htbp}{loh}
  \newfloat{scheme}{htbp}{los}

  \usepackage{graphicx}    

  \usepackage[english]{babel}
  \usepackage{array}

  \usepackage{subcaption}  
  \usepackage{outlines}    
  \usepackage[normalem]{ulem} 

  \usepackage[numbers,sort&compress,super]{natbib}
  \usepackage{natmove}

  \usepackage{cancel}      
  \usepackage{pdfpages}  
  \usepackage[font=scriptsize,labelfont=bf]{caption}

  \usepackage{hyperref} 
  \usepackage{cleveref}	
  	\crefname{figure}{Figure}{Figures}
  	\crefname{table}{Table}{Tables}
  	\crefname{equation}{Eq.}{Eqs.}
  	\crefname{section}{Section}{Sections}
  	\crefname{subsection}{Section}{Sections}
  	\crefname{subsubsection}{Section}{Sections}
  	\crefname{algorithm}{Algorithm}{Algorithms}

  \usepackage{tikz}

    \usepackage[ruled,vlined]{algorithm2e}

  \usepackage{todonotes}

\usepackage{epstopdf}
\AppendGraphicsExtensions{.tiff}

  \author{Karl Pierce}
  \affiliation{ Center for Computational Quantum Physics, \\Flatiron Institute, 162 5th Ave., New York, 10010  NY,  USA}
  \email{kpierce@flatironinstute.org}

  \title{Towards Using Matrix-Free Tensor Decompositions to Systematically Improve Approximate Tensor-Networks.}

\begin{document}

\begin{abstract}
We investigate a novel approach to approximate tensor-network contraction via the exact, matrix-free decomposition of full tensor-networks.
We study this method as a means to eliminate the propagation of error in the approximation of tensor-networks.
Importantly, this decomposition-based approach is generic, i.e. it does not depend on a specific tensor-network, the tensor index (physical) ordering, or the choice of tensor decomposition. 
Careful consideration should be made to determine the best decomposition strategy.
Furthermore, this method does not rely on robust cancellation of errors (i.e. the Taylor expansion).
As a means to study the effectiveness of the approach, we replace the exact contraction of the particle particle ladder (PPL) tensor diagram in the popular coupled-cluster with single and double excitation (CCSD) method with a low-rank tensor decomposition, namely the canonical polyadic decomposition (CPD).
With this approach, we replace an $\mathcal{O}(N^6)$ tensor contractions with a potentially reduced-scaling $\mathcal{O}(N^4R)$ optimization problem, where $R$ is the CP rank, and we reduce the computational storage of the PPL tensor from $\mathcal{O}(N^4)$ to $\mathcal{O}(NR)$, although we do not take advantage of this compression in this study.
To minimize the cost of the CPD optimization, we utilize the iterative structure of CCSD to efficiently initialize the CPD optimization.
We show that accurate chemically-relevant energy values can be computed with an error of less than 1 kcal/mol using a relatively low CP rank.
\end{abstract}

 \section{Introduction}
Multi-index arrays, also known as tensors, are an integral tool in the translation of mathematical models into computer algorithms. 
Unfortunately, tensors are plagued by the so-called {\it curse of dimensionality} which dictates that the computational effort required to access and manipulate data in a tensor grows exponentially with the dimensionality of the tensor.
Physics-based modeling algorithms, such as those found in electronic structure methods, are comprised of sets of tensor-tensor interactions commonly known as {\it tensor-networks}.
In an effort to overcome the curse of dimensionality associated with the evaluation of these tensor-networks, researchers often find compressed or low-rank representations of tensor components in these networks.
However, with each approximated component-tensor introduced into a tensor-network, a degree of error associated with that component is also introduced into the network.
One can express the error in a component-wise approximated tensor-network as the error of each approximated component propagated through the tensor-network.
The total accumulated error in an approximated tensor-network is therefore proportional to many variables such as total length of the tensor-network, total number of approximated tensor, error per approximated-tensor, etc.
To mitigate this error propagation, researchers have found some success in using the Taylor series expansion to remove leading-order error in component-wise approximated tensor-networks.\cite{VRG:dunlap:2000:PCCPP, VRG:pierce:2021:JCTC} 
However, this {\em robust} approximation approach cannot practically be applied to every approximated tensor-network.

In electronic structure methods there exists an incredibly large and diverse number of studied low-rank tensor approximations.
These approximations can be organized into two categories: data-sparse methods and rank-sparse methods.
Data-sparse methods (such as localization techniques like the pair-natural orbital approach,\cite{VRG:pulay:1983:CPL,VRG:pulay:1984:JCP,VRG:ahlrichs:1975:JCP,VRG:neese:2009:JCPa,VRG:neese:2011:JCTC,Rolik:2011:JCP,Rolik:2013:JCP} and fragmentation schemes\cite{VRG:kitaura:1999:CPL,Kristensen:2011:JCTC,Li:2002:JCC,Li:2006:JCP,Li:2009:JCP,VRG:guo:2018:JCP}) recast a higher-order tensor into a form which stores only non-zero valued elements of the tensor.
Rank-sparse approaches (such as the density fitting approximation\cite{VRG:whitten:1973:JCP,VRG:dunlap:1979:JCP,Vahtras:1993:CPL,VRG:jung:2005:PNAS,VRG:mintmire:1982:PRA}) recast higher-order tensors into compact tensor-network-based decompositions by revealing latent structures of the tensor.
In principle, these techniques are integrated into electronic structure methods by first identifying high-cost tensors-networks and subsequently replacing single or multiple components of these networks with low-rank approximations.
While there are many successful approximations introduced across the domain, here we investigate an alternative means to approximate a tensor-network with directly controllable accuracy, as opposed to the indirect control associated with component-wise approximations.

The direction of this work is to extend the application of our previous work\cite{VRG:pierce:2023:JCTC,Pierce:2025:JCTC} to approximate any high-cost tensor-network, not just the two-electron integral tensor.
By construction, optimizing the low-rank decomposition of a tensor-network provides a direct means to control error.
Importantly, with this approach we replace the cost of contracting a tensor-network with the cost of optimizing the low-rank decomposition and, furthermore, we replace the computational storage cost associated with the contracted tensor network with the cost associated with the chosen low-rank decomposition format. 
Although we should note, we do not take advantage of this storage benefit in this preliminary study.
With these properties, we believe that this scheme is an innovative means to explore novel low-scaling formulations of electronic structure methods.
To make our idea more concrete, in this work we construct the particle-particle ladder (PPL) diagram in the coupled cluster with single and double excitations (CCSD) method using a matrix-free canonical polyadic decomposition (CPD).
We emphasize that this approach is a general technique that is not limited to this specific tensor-network, the coupled cluster method or the use of the CPD.
It should also be noted that the focus of this study is to demonstrate the accurate application of matrix-free optimizations to electronic structure methods and we do not seek to reduce the wall-time of the CCSD method.

The rest of this manuscript is organized as follows. In \cref{sec:formalism} we introduce the coupled cluster method's particle-particle ladder (PPL) tensor-network, demonstrate how state of the art approximations to this network introduce unintended error and propose a reduced-scaling, matrix-free CPD approximation of the PPL tensor-network.
In \cref{sec:compdetails} we discuss the details of our computational experiments. 
In \cref{sec:results} we report on the capacity for a matrix-free CPD optimization to replace the exact PPL tensor-network. 
Finally, \cref{sec:conclusions} summarizes our findings and provides future research directions for efficient application of this tensor-network approximation schemes to this and other threads of research.

\section{Formalism}
 \label{sec:formalism}

\subsection{Coupled Cluster and the Particle-Particle Ladder Network}
In this work we demonstrate how the decomposition of an entire tensor-network can replace the exact contraction of the tensor-network with controllable accuracy.
With this approach we are provided a level of error control that is missing from today's standard methods for approximating tensor-networks.
In this preliminary study, we apply this technique to a representative tensor-network from the popular CCSD excitation method.
In this section we introduce the high-cost tensor-network from the CCSD method and demonstrate how the state-of-the-art tensor hyper-contraction approach introduces error into the approximation of the tensor-network.
Please note, an in-depth description of the CC optimization will not be discussed in this work and readers can find extensive reviews on the topic, such as Ref. \citenum{VRG:crawford:2000:RCC}.
In \cref{sec:CPD-PPL}, we propose a method to approximate the tensor-network with reduced-scaling and controllable accuracy using a matrix-free tensor decomposition optimization strategy.

The goal of the coupled cluster method is to recover the ‘dynamical correlation’ that is missing from the Slater determinant picture of Hartree-Fock Theory.\cite{VRG:cizek:1966:JCP, Cizek:1969:ACP,Cizek:1971:IJQC,VRG:crawford:2000:RCC,VRG:tajti:2004:JCP,VRG:harding:2008:JCP,Thorpe:2019:JCP,VRG:purvis:1982:JCP}
The CCSD method achieves this goal by optimizing the weights of the single and double excitation amplitude tensors using a coupled set of nonlinear equations, i.e.
\begin{align}\label{eq:ccsd}
    E_\mathrm{CC} &= \langle \Phi_0 | \bar{H} | \Phi_0 \rangle \\ \nonumber
    0 &= \langle \Phi^{a}_{i} | \bar{H} | \Phi_0 \rangle \\ \nonumber
    0 &= \langle \Phi^{ab}_{ij} | \bar{H} | \Phi_0 \rangle
\end{align}
where $\bar{H}$ is the similarity transformed Hamiltonian and $E_\mathrm{CC}$ is the coupled cluster energy.
The CCSD similarity transformed Hamiltonian can be expressed as $\bar{H} = e^{-\hat{T}_\mathrm{SD}} H e^{\hat{T}_\mathrm{SD}}$ where $\hat{T}_\mathrm{SD}$ is the excitation operator truncated at double excitations:
\begin{align}
    \hat{T}_\mathrm{SD}= \hat{T}_1 + \hat{T}_2.
\end{align}
The restricted, closed-shell single and double excitation operators can be expressed as 
\begin{align}
    \hat{T}_1 &= t^a_i \hat{E}^i_a \\ \nonumber
    \hat{T}_2 &= t^{ab}_{ij} \hat{E}^{ij}_{ab}
\end{align}
where $t^a_i$ is the singlet excitation amplitude tensor and $\hat{E}^i_a = a^\dagger_\alpha i_\alpha + a^\dagger_\beta i_\beta$ is the singlet excitation operator expressed as a sum of products of creation and annihilation operators.
Similar definitions can be written for the $t^{ab}_{ij}$ and $\hat{E}^{ij}_{ab}$.
In \cref{eq:ccsd}, the labels $i,j,k,...$ are representative of the set of $O$ occupied orbital basis functions and the labels $a,b,c,...$ are representative of the set of $V$ unoccupied orbital basis functions.
The weights, $t^a_i$ and $t^{ab}_{ij}$, are then determined through the optimization of \cref{eq:ccsd}.
Although the Coupled Cluster family of methods find great success in recovering dynamical correlation, the family is plagued by a restrictively large computational complexity.
\begin{figure}[t]
    \includegraphics[clip, trim=5cm 16.5cm 5cm 8cm,width=0.5\textwidth]{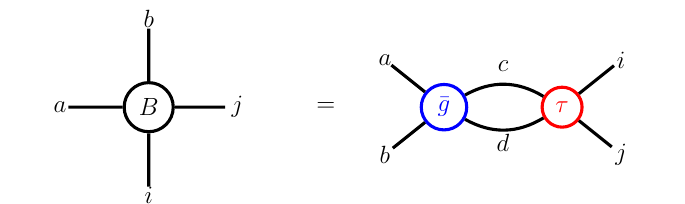}
    \caption{Diagrammatic representation of the particle-particle ladder tensor-network.}
    \label{fig:PPL-TN}
\end{figure}

The complexity of CCSD comes from the evaluation of tensor-networks in the residual equations of the nonlinear optimization of $t^a_i$ and $t^{ab}_{ij}$.
We introduce the most expensive term in the CCSD residual equations: the particle-particle ladder (PPL) tensor-network.
In restricted, closed-shell format, the residual symmetrized PPL tensor-network can be written as
\begin{align}\label{eq:B}
    B^{ab}_{ij} = \sum_{cd} (g^{ab}_{cd} - 2 \sum_{k} g^{ab}_{ck} t^{k}_{d}) \tau^{cd}_{ij} 
\end{align}
where $\tau^{cd}_{ij} = t^{cd}_{ij} + t^{c}_{i}t^{d}_{j}$ and $g$ is the two-electron integral (TEI) tensor that will be discussed in \cref{sec:df_sec}. 
As an aside, \cref{eq:B} can be constructed using the intermediate $\bar{g}$ such that
\begin{align}\label{eq:gbar}
    \bar{g}^{ab}_{cd} = g^{ab}_{cd} - 2 \sum_k g^{ab}_{ck} t^k_d
\end{align}
resulting in simplified expression
\begin{align}\label{eq:ppl}
    B^{ab}_{ij} = \sum_{cd} \bar{g}^{ab}_{cd} \tau^{cd}_{ij}.
\end{align}
\cref{fig:PPL-TN} shows the diagrammatic expression of the PPL tensor-network (\cref{eq:ppl}). 
In this figure, tensors are represented as nodes on a graph and tensor indices are edges. A tensor contraction (or summation over an index) is denoted as a solid line connecting two nodes and, later, a hyperedge (an index which connects more than two nodes) will be denoted as a dashed line.
Also, the {\em order} of a tensor is defined as the number of indices or edges which belong to a tensor.
From either of these representations, one can see that the computational complexity of computing the PPL tensor-network is $O^2V^4 \approx \mathcal{O}(N^6)$ where $N$ is a measure of the system-size.

Although there are many formally $\mathcal{O}(N^6)$ terms in the CCSD residual equation, the PPL tensor-network is the only term that carries an $O^2V^4$ cost of evaluation.
Because of the PPL tensor-network's dominating cost and the prevalence of previous studies which have approximated the PPL tensor-network using component-wise decomposition schemes,\cite{VRG:pierce:2021:JCTC,VRG:hummel:2017:JCP,VRG:parrish:2014:JCP,Datar:2024:JCTC,Martinez:1995:JCP,Martinez:1993:JCP} we choose to apply our decomposition strategy to only this term in CCSD.
Please note that our goal is not to eliminate the error associated with approximating the PPL tensor network, but to demonstrate that it is possible to replace a tensor-network contraction with a low-rank tensor decomposition.
Therefore, error will still exist and propagate into the optimization of the wavefunction excitation tensors.
Although, we believe that the analysis and mitigation of this error propagation should be considered in future studies.
Below, we will introduce two canonical decomposition schemes and demonstrate how they can introduce a degree of error propagation in the evaluation of the PPL tensor-network.

\subsubsection{The Density Fitting approximation}\label{sec:df_sec}
The most standard approximation made in electronic structure methods is the density fitting (DF) approximation\cite{VRG:whitten:1973:JCP,VRG:dunlap:1979:JCP,Vahtras:1993:CPL,VRG:jung:2005:PNAS}  of the TEI tensor, $g$. 
The elements of the TEI tensor evaluated in a generic set of basis functions $\{ \phi \}$ are
\begin{align}\label{eq:g}
    g^{pq}_{st} = \iint \phi^*_{p}(r_1) \phi^*_q(r_2) g(r_1, r_2) \phi_s(r_1) \phi_t(r_2) dr_1 dr_2.
\end{align}
In general, $g(r_1, r_2)$ may be any positive kernel and for this work we utilize the Coulomb interaction kernel, i.e. $g(r_1, r_2) \equiv | r_1 - r_2 |^{-1}$.
The DF approximation is a low-rank representation of the TEI tensor defined as
\begin{align}\label{eq:SQG}
    g^{pq}_{st} \overset{\mathrm{DF}}{\approx} \sum_X D^{pX}_{s} D^{qX}_{t}
\end{align}
where the index $X$ here represents an optimized, predetermined auxiliary set of basis functions which grows linearly with system size and $D$ is denoted as the DF approximated TEI tensor.
It should be noted that the DF approximation can alternatively be constructed analytically using the Cholesky decomposition\cite{VRG:beebe:1977:IJQC,VRG:lowdin:1965:JMP,Lowdin:2009:IJQC,Folkestad:2019:JCP} or the  related chain-of-spheres (COSX) method.\cite{VRG:izsak:2011:JCP,Izsak:2012:MP,VRG:dutta:2016:JCP,VRG:izsak:2013:JCP,VRG:neese:2009:CP,Kossmann:2010:JCTC,Kossmann:2009:CPL}
One can make \cref{eq:SQG} exact by introducing a DF error correction term $(\Delta g)^{ps}_{qt}$ into the equation, i.e.
\begin{align}
        g^{pq}_{st} = \sum_X D^{pX}_{s} D^{qX}_{t} + (\Delta g)^{pq}_{st}.
\end{align}
Because of the significant compression gained from the DF approximation, the error associated with replacing the TEI tensor with its DF approximation is widely accepted and this work is no exception.\cite{VRG:weigend:2002:JCP, Hattig:2005:PCCP,VRG:deprince:2013:JCTC}
By this we assume $(\Delta g)^{pq}_{st} \approx 0$ for all DF approximated TEI tensors and in CCSD, we will replace all TEI tensors with their DF approximations, accordingly.

\begin{figure}[t]
    \includegraphics[clip, trim=5cm 16.5cm 5cm 8cm,width=0.5\textwidth]{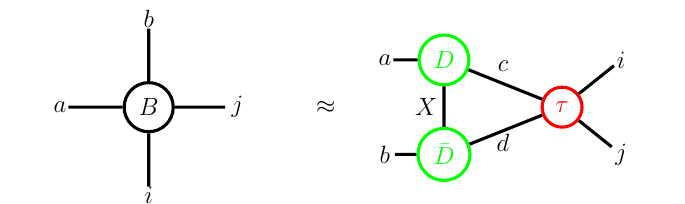}
    \caption{Diagrammatic representation of the DF approximated PPL tensor-network}
    \label{fig:DF-PPL-TN}
\end{figure}

In principle, the goal of introducing the DF approximation is to reduce the computational complexity of the tensor-networks in the CCSD method.
In practice, the DF approximation fails to reduce the complexity of {\bf all} tensor-networks in the CCSD method.
Specifically, it is unable to reduce the complexity of the PPL tensor-network.
By substituting the appropriate DF approximated TEI tensors into \cref{eq:B} we find the following expression
\begin{align}\label{eq:fulldfB}
    B^{ab}_{ij} = \sum_{cd} \sum_X(D^{aX}_{c} D^{bX}_d - 2 \sum_{k} D^{aX}_{c} D^{bX}_{k} t^{k}_{d}) \tau^{cd}_{ij}.
\end{align}
To simplify this expression, it is possible to form an intermediate $\bar{D}$ term which is similar to $\bar{g}$ from \cref{eq:gbar},
\begin{align}
    \bar{D}^{bX}_{d} = (D^{bX}_{d} - 2 \sum_{k} D^{b}_{k} t^{k}_{d}).
\end{align}
Substituting $\bar{D}$ into \cref{eq:fulldfB} simplifies the expression to be
\begin{align}\label{eq:dfB}
    B^{ab}_{ij} = \sum_{cd} (\sum_{X} D^{aX}_{c} \bar{D}^{bX}_{d})\tau^{cd}_{ij}.
\end{align}
We express the diagrammatic representation of this network in \cref{fig:DF-PPL-TN} and this figure allows one to easily see that the indices $c$ and $d$ occupy different order-3 $D$ tensors.
By construction, the DF approximation pins pairs of orbital indices of a single electronic particle to the same $D$ tensor and, therefore, this observation means that $c$ and $d$ are indices of different particles.
The DF approximated PPL tensor-network does not have a reduced complexity over the canonical PPL tensor-network.

Because the computation of the PPL tensor-network is a significant bottleneck in the CCSD optimization procedure, theorists have studied additional approximations that can effectively reduce the scaling of this tensor-network.
These include the pseudospectral\cite{VRG:friesner:1985:CPL,Friesner:1986:JCP,Langlois:1990:JCP,VRG:ringnalda:1990:JCP,Friesner:1991:ARPC,Martinez:1995:JCP,VRG:martinez:1992:JCP,Martinez:1994:JCP,Ko:2008:JCP,Martinez:1993:JCP} and the tensor hyper-contraction\cite{VRG:hohenstein:2012:JCP,VRG:hohenstein:2012:JCPa,VRG:parrish:2012:JCP,Hohenstein:2013:JCP,VRG:parrish:2014:JCP,Shenvi:2013:JCP,Schutski:2017:JCP,Parrish:2019:JCP,Lee:2019:JCTC,VRG:hummel:2017:JCP,VRG:song::JCP,Hohenstein:2019:JCP,Hohenstein:2021:JCP,Hohenstein:2022:JCP,Jiang:2022:JCTC,Zhao:2023:JCTC,Datar:2024:JCTC,Schmitz:2017:JCP, Khoromskaia:2015:PCCP} methods.
In the following section, we will discuss how the popular tensor hyper-contraction is formulated and how the approximation introduces error that is propagated through the PPL tensor-network.

\subsubsection{Propagation of error in the THC approximation}
As shown in the previous section, the DF approximation cannot reduce the computational scaling of the expensive PPL tensor-network.
It is well understood that the tensor hyper-contraction (THC) method\cite{Hohenstein:2013:JCP,VRG:parrish:2014:JCP,Shenvi:2013:JCP,Schutski:2017:JCP} can be deployed to reduce the complexity of this tensor-network.
Therefore, in this section we introduce the THC approximation and highlight the error it contributes to the construction of the PPL tensor-network.
Please note,  we choose this example because its behavior has been studied and documented\cite{VRG:pierce:2021:JCTC} and not because we believe that the THC method does a poor job approximating the PPL tensor-network or because this error propagation issue is unique to the THC approximation.

\begin{figure}[t]\label{fig:THC-PPL-TN}
\begin{subfigure}{0.49\textwidth}
        \includegraphics[clip, trim=5cm 16.5cm 5cm 8cm,scale=0.8]{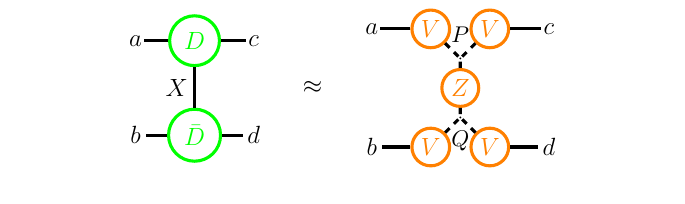}
    \caption{}
    \label{fig:THC-TN}
    \end{subfigure}
    \begin{subfigure}{0.5\textwidth}
        \includegraphics[clip, trim=5cm 16.5cm 5cm 8cm, scale=0.8]{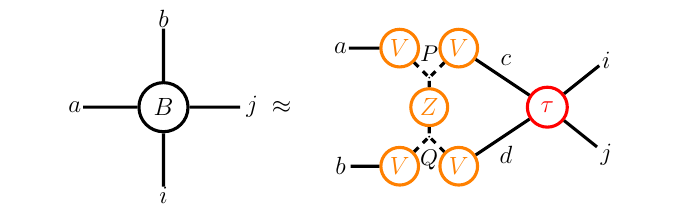}
        \caption{}
    \label{fig:THC-PPL}
    \end{subfigure}\hfill
    \caption{a) Diagrammatic representation of the tensor hyper-contraction approximation b)Diagrammatic representation of the THC approximation of the PPL tensor-network}
\end{figure}

The THC is a low-rank decomposition of the order-4 TEI tensor such that
\begin{align}\label{eq:fullTHC}
    g^{ab}_{cd} \approx \sum_{PQ} V^a_P V^P_c Z^P_Q V^{b}_Q V^Q_d.
\end{align}
The THC representation of the DF approximated TEI can be visualized using \cref{fig:THC-TN}.
In principle, the THC can be formed by replacing both order-3 DF approximated TEI tensors with the low-rank approximation,
\begin{align}\label{eq:thcd_sum}
    D^{pX}_{q} &\approx \sum_P V^{p}_{P} V^{P}_{q} V^P_X
\end{align}
where the THC rank, $P$, can be determined analytically using a rank revealing CPD or using a predefined grid basis set. 
In \cref{eq:thcd_sum}, the $V$ tensors are denoted as THC factor matrices.
In this example, the THC core matrix is defined as $Z^P_Q = \sum_{X} V^{P}_{X} V^{X}_{Q}$.
However, $Z$ can also be computed using alternative methods, such as a low-cost least-squares optimization\cite{VRG:parrish:2012:JCP} which will not be discussed in this work.

We can make the approximation in \cref{eq:thcd_sum} exact by introducing a THC approximation error term $\Delta D$:
\begin{align}
    D^{pX}_{q} &= \sum_P V^{p}_{P} V^{P}_{q} V^P_X + (\Delta D)^{pX}_q.
\end{align}
And for simplicity in the following analysis we will define
\begin{align}
    \tilde{D}^{pX}_q = \sum_P V^{p}_{P} V^{P}_{q} V^P_X
\end{align}
where a tilde indicates the THC approximation of a given tensor; therefore
\begin{align}\label{eq:thcD}
     D^{pX}_{q} &= \tilde{D}^{pX}_q + (\Delta D)^{pX}_q.
\end{align}
If we substitute \cref{eq:thcD} into the definition of the DF approximation of the TEI tensor, \cref{eq:SQG}, we find the following
\begin{align}\label{eq:errorTHC}
    g^{ab}_{cd} = \sum_X (\tilde{D} + \Delta D)^{aX}_{c} (\tilde{D} + \Delta D)^{bX}_{d}.
\end{align}
\cref{eq:errorTHC} can be expanded into 
\begin{align}\label{eq:full_thc_error}
     g^{ab}_{cd} = \sum_X \tilde{D}^{aX}_{c} \tilde{D}^{bX}_d + (\Delta D)^{aX}_{c}  \tilde{D}^{bX}_d 
     + \tilde{D}^{aX}_{c} (\Delta D)^{bX}_{d} + (\Delta D)^{aX}_{c} (\Delta D)^{bX}_{d}
\end{align}
and, as one can see from this expression, the accuracy of the THC approximated TEI tensor depends on the error associated with the THC approximation propagated through \cref{eq:SQG}.
Similar to \cref{eq:errorTHC}, we can define the error in the THC approximation of $g$ as 
\begin{align}
    (\Delta g)^{ab}_{cd} = \sum_X (\Delta D)^{aX}_{c}  \tilde{D}^{bX}_d + \sum_X \tilde{D}^{aX}_{c} (\Delta D)^{bX}_{d} + \sum_X (\Delta D)^{aX}_c (\Delta D)^{bX}_d).
\end{align}
This error term simplifies \cref{eq:full_thc_error} into the following expression
\begin{align}\label{eq:simplifiedTHC}
    g^{ab}_{cd} = \tilde{g}^{ab}_{cd} + (\Delta g)^{ab}_{cd}
\end{align}
where $\tilde{g}^{ab}_{cd} \equiv \sum_{PQ}V^a_P V^P_c Z^P_Q V^{b}_Q V^Q_d$ is the THC approximation of $g$.

Introducing \cref{eq:simplifiedTHC} into the PPL tensor-network results in the following expression
\begin{align}
    B^{ab}_{ij} &= \sum_{cd} (\tilde{g}^{ab}_{cd} + (\Delta g)^{ab}_{cd}) \tau^{cd}_{ij} \\ \nonumber
    &= \sum_{cd} \tilde{g}^{ab}_{cd}\tau^{cd}_{ij} + (\Delta B)^{ab}_{ij}
\end{align}
where $(\Delta B)^{ab}_{ij}$ is the error in the THC approximated PPL tensor-network.
As an aside, the THC approximated PPL tensor-network is represented in \cref{fig:THC-PPL} and with this figure, it is easy to see how the THC approximation can be used to reduce the complexity of the PPL tensor-network.
The total, accumulated error in the THC approximated PPL tensor-network can be expanded as
\begin{align}\label{eq:error_ppl_thc}
    \Delta B^{ab}_{ij} &= \sum_{cd}(\Delta g)^{ab}_{cd} \tau^{cd}_{ij} \\ \nonumber
    &= \sum_{Xcd} (\Delta D)^{aX}_{c}  \tilde{D}^{bX}_d \tau^{cd}_{ij} +  \sum_{Xcd} \tilde{D}^{aX}_{c} (\Delta D)^{bX}_{d} \tau^{cd}_{ij} +  \sum_{Xcd} (\Delta D)^{aX}_c (\Delta D)^{bX}_d \tau^{cd}_{ij}.
\end{align}
This equation shows that the error in the PPL tensor-network is a sum of error tensors propagated through the PPL tensor network.
In our previous work,\cite{VRG:pierce:2021:JCTC} we introduced the robust approximation of the THC approximated TEI tensor. 
This approximation uses a Taylor series expansion to identify terms which can be added to a component-wise approximated tensor network to cancel the leading-order error propagation.
The error in the robust THC approximated PPL tensor-network can be written as
\begin{align}
    \Delta B^{ab}_{ij} &\overset{rCP}{=} \sum_{Xcd} (\Delta D)^{aX}_c (\Delta D)^{bX}_d \tau^{cd}_{ij}.
\end{align}
Although significantly suppressed, this equation shows that the error in the robust component-wise approximated PPL tensor-network is still a result of error-propagation, i.e. it cannot be determined by the error in the THC approximation alone.
Furthermore, although the robust approximation was effective at mitigating error propagation for this network, it is not easily applicable to all tensor-networks.

In the remainder of this work, we introduce a method to approximately contract the PPL tensor-network in a way that makes the error in the contraction a directly controllable component of our approximation.
To accomplish this task, we compute the canonical polyadic decomposition of the entire DF approximated PPL tensor-network.
In the following section, we remind the readers on the details of the canonical polyadic decomposition before using it to approximate the PPL tensor-network.

\begin{figure}
    \includegraphics[clip, trim=5cm 16.3cm 5cm 8cm,width=0.5\textwidth]{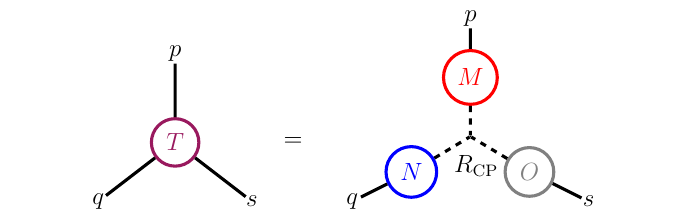}
    \caption{Diagrammatic representation of the canonical polyadic decomposition of a tensor $T$.}
    \label{fig:CPD}
\end{figure}
\subsection{The Canonical Polyadic Decomposition}
The CANDECOMP/PARAFAC, or, simply, the canonical polyadic decomposition (CPD)\cite{VRG:carroll:1970:P,VRG:Harshman:1970:WPP} is a data-sparse representation of higher-order tensors that has been used with various success in computational modeling to reduce the cost and/or complexity of the tensor algebra which accompanies physical simulation.
We discuss various applications of the CPD to electronic structure methods in \cref{sec:lit_rev}.
The CPD of an order-$n$ tensor can be understood as a compression of a tensor of $n$ modes to a tensor-network of $n$ order-$2$ tensors connected by a single hyperedge.
For example, given the tensor $T \in \mathbb{R}^{I_P \times I_Q \times I_S}$, the CPD can be expressed as
\begin{align}
  T_{pqs} \overset{\mathrm{CPD}}{=} \overset{\approx}{T}_{pqs} = \sum_{r}^{R_\mathrm{CP}} \lambda_r M_{pr} N_{qr} O_{sr}
\end{align} 
where $[M, N, O]$, are the set of CP factor matrices and $R_{\mathrm{CP}}$ is the CP rank of the decomposition.
We enforce that the factor matrices be column-wise normalized and store the column-wise scaling factors in $\lambda$.
Without loss of generality, it is possible to factorize lambda into any one or many of the factor matrices and, therefore, we will ignore it in future CP-based equations. 
Notice, too, that we denote the CP approximation of a tensor with a double tilde.
The diagrammatic representation of the CPD can be found in \cref{fig:CPD} where the dashed line represents the CP hyperedge index.

As one can see, the topology of the CPD is relatively simple and the effectiveness of the CP compression depends explicitly on the extent of the CP hyperedge (CP rank).\cite{Hastad:1990:algorithm, VRG:hillar:2013:JA} 
Unfortunately, there exists no simple formula that can determine the CP rank.
Therefore, the value is revealed by constructing multiple rank-$R$ CPD approximations and choosing the value of $R$ which satisfies the problem's predetermined accuracy thresholds.
Therefore, a more accurate representation of the CPD would be
\begin{align}
    T_{pqs} = \sum_{r}^{R} M_{pr} N_{qr} O_{sr} + (\Delta T)_{pqs}
\end{align}
where $\Delta T$ is the error in the CPD approximation.
In the following sections we demonstrate how we utilize the CPD to approximate the PPL tensor-network in CCSD.

\begin{figure}[b]
    \includegraphics[clip, trim=5cm 15.2cm 5cm 8cm,width=0.5\textwidth]{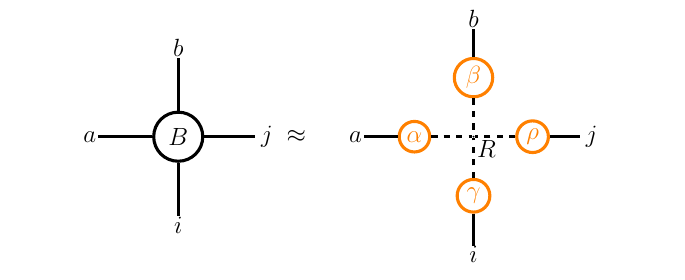}
    \caption{Diagrammatic representation of CPD approximation of the PPL tensor}
    \label{fig:CP-PPL-TN}
\end{figure}

\subsection{The CPD approximation of the PPL tensor-network}\label{sec:CPD-PPL}
Our goal is to develop a method to approximate tensor-networks such that the computational complexity of the tensor-network is reduced and the error in the network’s approximation is exactly controllable through the optimization process.
In this example, we achieve this goal by representing the DF approximated PPL tensor-network (\cref{eq:dfB}) in a low-rank CPD format:
\begin{align} \label{eq:cp-ppl}
    B^{ab}_{ij} \overset{\mathrm{CP-PPL}}{\approx} \sum_{r}^{R}\alpha^a_r \beta^b_r \gamma^r_{I} \rho_j^r,
\end{align}
in which we denote the CP-PPL approximation.
In this equation $R$ is the rank of the decomposition and $\alpha, \beta, \gamma$ and $\rho$, are the CP factor matrices.
A diagrammatic representation of the CP-PPL approximation is shown in \cref{fig:CP-PPL-TN}.

Similar to the previous approximations, the CP-PPL can be made exact by introducing the CP-PPL error tensor $\Delta B$.
For convenience we define the CPD approximation 
\begin{align}
    \overset{\approx}{B}{}^{ab}_{ij} = \sum_r^R \alpha^a_r \beta^b_r \gamma^r_{i} \rho_j^r
\end{align}
and the exact decomposition as
\begin{align}
    B^{ab}_{ij} = \overset{\approx}{B}{}^{ab}_{ij} + \Delta B^{ab}_{ij}.
\end{align}
With this approximation the error in the CP-PPL is correlated with the compressibility of the tensor $B$ and the ability to find accurate CP factor matrices for a given rank.

We construct an accurate rank $R$ CP-PPL approximation in a {\em matrix-free} way by minimizing the following loss function 
\begin{align}\label{eq:loss_function}
  f(\overset{\approx}{B}) = \frac{1}{2} \| \sum_{cd} (\sum_{X}D^{aX}_{c} \bar{D}^{bX}_{d}) \tau^{cd}_{ij} - \overset{\approx}{B}{}^{ab}_{ij} \|^2_2.
\end{align}
The error in the approximated CP-PPL tensor-network is then exactly the evaluation of the function $f(\overset{\approx}{B})$.

In principle this does not mean that a given fixed rank CP-PPL decomposition will be a {\bf better} approximation than a given component-wise approximation of the network, just that the error in the approximate contraction of the  tensor-network is controllable through the minimization of $f$.
The optimization of $\overset{\approx}{B}$ is called matrix-free 
because we effectively "load" the output of the PPL network into the CPD factor matrices without a-priori computing the tensor $B$.
To follow, we attempt to replace the exact PPL tensor-network with the CP-PPL optimization, \cref{eq:loss_function}.
It should be noted that, in this study, $\overset{\approx}{B}$ is not preserved in the CPD form after a rank $R$ optimization is completed. Instead the tensor is reconstructed into its canonical form for the remainder of the CCSD optimization algorithm.
We also point out that we do not explicitly enforce symmetry in our minimization of \cref{eq:loss_function}, i.e. $\overset{\approx}{B}{}^{ab}_{ij} \neq \overset{\approx}{B}{}^{ba}_{ji}$.
But, because we do not preserve the low-rank structure of $B$, we recover this symmetry in the post tensor-reconstruction by defining
\begin{align}
    \overset{\approx}{B}{}^{ab}_{ij} = \frac{1}{2} (\overset{\approx}{B}{}^{ab}_{ij} + \overset{\approx}{B}{}^{ba}_{ji}).
\end{align}
For a given CP rank we minimize \cref{eq:loss_function} using a simple alternating least squares (ALS) optimization\cite{VRG:kroonenberg:1980:P,VRG:beylkin:2002:PNAS} and in the following section we evaluate the computational scaling of each step in the CP-PPL ALS optimization.

\subsection{Optimizing the CP-PPL tensor-network}
In this section, we walk through the computational cost associated with optimizing \cref{eq:loss_function} using an ALS algorithm. 
In doing so, we demonstrate that such an optimization has a reduced-scaling compared to the canonical evaluation of the PPL tensor-network.
In general, computing a CP-ALS update involves four steps.
First, one chooses a CPD factor matrix in the cost function, for example the alpha from the $n$th iteration ($\alpha_n$), and fixes the other factors.
Second, one takes the gradient of the CP loss function, here \cref{eq:loss_function}, with respect to the chosen factor matrix. 
The gradient of $f(\overset{\approx}{B}{}_n)$ with respect to $\alpha_n$ is 
\begin{align}
    \frac{\partial f(\overset{\approx}{B}{}_n)}{\partial \alpha_n} &= \frac{1}{2}\frac{\partial}{\partial \alpha_{n}}(B^2 - B \overset{\approx}{B}{}_n - \overset{\approx}{B}{}_nB +\overset{\approx}{B}{}^2_n) \\ \nonumber
    &= -B\frac{\partial \overset{\approx}{B}{}_n}{\partial \alpha_n} + \overset{\approx}{B}{}_n \frac{\partial\overset{\approx}{B}{}_n}{\partial \alpha_n}
\end{align}
where $\overset{\approx}{B}{}_n$ is the CPD approximation of $B$ during at the $n$th ALS iteration and 
\begin{align}
    (\frac{\partial \overset{\approx}{B}{}_n}{\partial \alpha_n})^{rb}_{ij} = (\beta_n)^b_r (\gamma_n)^r_i (\rho_n)^r_j.
\end{align}
Next, one sets the gradient expression from the previous step equal to zero and solves the developed linear problem to determine a new optimal factor matrix, $\alpha_{n+1}$.
This process can be repeated for each of the problem’s other factor matrices and sweeps of factor matrix updates can be repeated as many times as necessary.
Finally, after a round of updating each factor matrix a stopping condition is assessed. 
We will discuss appropriate stopping conditions in \cref{sec:cpdstop}. 
Next, we evaluate the cost of computing the CP-PPL gradient and finding an updated CP factor matrix.

To exemplify the computational cost of the optimization, we walk through the optimization of the $\alpha_{n+1}$ factor matrix.
Please note, in this analysis we assume that the CP rank grows linearly with system size based on findings compiled in \cref{sec:results} and, for simplicity, we drop the ALS iteration label ($n$).
The $\alpha_{n+1}$ solution to the ALS update for the CP-PPL cost function is
\begin{align}\label{eq:full_der}
 [\alpha_{n+1}]^{a}_{r^\prime} = \sum_{cdbijr}(\sum_{X}D^{aX}_{c} \bar{D}^{bX}_{d}) t^{cd}_{ij} \beta^b_r \gamma^r_i \rho^r_j [(\sum_b \beta^b_r \beta^b_{r^\prime}) (\sum_i \gamma^r_i \gamma^{r^\prime}_i) (\sum_j \rho^r_j \rho^{r^\prime}_j)]^{-1}.
\end{align}
In the literature, it is common to group all of the terms in the square bracket into a term denoted as $W^{r}_{r^\prime}$.
\begin{align}
    W^r_{r^\prime} = (\sum_b \beta^b_r \beta^b_{r^\prime}) (\sum_i \gamma^r_i \gamma^{r^\prime}_i) (\sum_j \rho^r_j \rho^{r^\prime}_j).
\end{align}
Computation of $W$ can be done efficiently with a cost of $3NR^2 + 2R^2 \approx \mathcal{O}(N^3)$ and its inversion can also be done conveniently using a symmetric, square linear solve algorithm such as the Cholesky decomposition.
The inversion of $W$ has a cost of $R^3 \approx \mathcal{O}(N^3)$

Substituting $W$ into  \cref{eq:full_der} simplifies the equation to
\begin{align}
  [\alpha_{n+1}]^{a}_{r^\prime} = \sum_r[\sum_{Xc} D^{aX}_{c} (\sum_{d}(\sum_b \bar{D}^{bX}_{d} \beta^b_r) (\sum_{ij} \gamma^{r}_{i} \rho^r_j \tau^{cd}_{ij}))][W^{-1}]^r_{r^\prime}
\end{align}
Next, one can compute the two innermost contractions $\tau^{cd}_{r} = \sum_{ij} \gamma^{r}_{i} \rho^r_j \tau^{cd}_{ij}$ and $\bar{D}^{rX}_{d} = \sum_b \bar{D}^{bX}_{d} \beta^b_r$. These contractions have a cost of $N^4R \approx \mathcal{O}(N^5)$ and $N^3R \approx \mathcal{O}(N^4)$, respectively.
The intermediates produced from these contractions can be cached to help minimize the cost of repeated gradient evaluations.
After these contractions the following equation remains
\begin{align}\label{eq:distribute}
  [\alpha_{n+1}]^{a}_{r^\prime} = \sum_r[\sum_{Xc} D^{aX}_{c} (\sum_{d}\bar{D}^{rX}_{d} \tau^{cd}_{r})][W^{-1}]^r_{r^\prime}.
\end{align}
The terms in the left set of square brackets can be efficiently computed for each value of the rank as
\begin{align}\label{eq:foreachr}
    \sum_{Xc} D^{aX}_{c} ( \sum_d {}^{r}\bar{D}^X_d \text{ }{}^r\tau^c_d)
\end{align}
where ${}^{r}\bar{D}^X_d$ and ${}^r\tau^c_d$ are matrices associated with a specific value of $r$.
In \cref{eq:foreachr}, the sum over $d$ can be computed efficiently using a standard level-3 BLAS {\it gemm} kernel and the sum over the compound index $\{Xc\}$ can be computed using a standard level-2 BLAS {\it gemv} kernel.
The cost of both of these operations is $N^3$ for each value of $r$, leading to an overall scaling of $N^3R \approx \mathcal{O}(N^4)$. 
It is obvious that contractions in \cref{eq:distribute,eq:foreachr} could be distributed over blocks of $r$ to improve computational performance.
Although, we plan to exploit this parallelism in future work as our current algorithm is implemented in a serial tensor algebra library.
Finally one is left with 
\begin{align}
    [\alpha_{n+1}]^{a}_{r^\prime} = \sum_r D^{a}_{r}[W^{-1}]^r_{r^\prime}
\end{align}
which can be computed with a cost of $NR^2 \approx \mathcal{O}(N^3)$.
As one can now see, the computational complexity of an ALS update is $\mathcal{O}(N^5)$, which is a reduction over the canonical complexity of the PPL tensor-network contraction. 
Readers can please refer to \cref{table:scaling} for a consolidated representation of the algorithmic cost.
In \cref{sec:results} we will show that an approximately linear-scaling and modest value of $R$ is necessary to compute chemically accurate energy differences.
\begin{table}
    \centering
    \begin{tabular}{|c|c|c|} \hline 
        Contraction &  Cost  & Scaling \\ \hline 
        $\sum_{ij} \gamma^{r}_{i} \rho^r_j \tau^{cd}_{ij}$ & $O^2V^2R$ & $N^5$ \\ \hline 
        $ \sum_b \bar{D}^{bX}_{d} \beta^b_r$ & $V^2XR$ & $N^4$\\ \hline 
        $\sum_d\bar{D}^{rX}_d \tau^{rc}_d$ & $V^2XR$ & $N^4$ \\ \hline 
        $\sum_{Xc} D^{aX}_{c} \tau^{r}_{Xc}$ & $V^2XR$ & $N^4$\\ \hline
        $\sum_b \beta^b_r \beta^b_{r^\prime}$ & $VR^2$ & $N^3$\\ \hline 
        $W^{-1}$ & $R^3$ & $N^3$ \\ \hline
        $\sum_{r}D^{a}_{r}[W^{-1}]^r_{r'}$ & $VR^2$ & $N^3$\\ \hline
    \end{tabular}
    \caption{Cost and Scaling of each contraction in the CP-PPL optimization.}
    \label{table:scaling}
\end{table}

We do concede that CPD optimization does have a large computational prefactor due in part to its hyperedge structure and in part to the necessary repeated evaluations of the gradient. 
In this work, we do our best to minimize the prefactor by caching intermediate tensors, implementing efficient kernels and developing an effective initialization, discussed in \cref{sec:cpdinit}.
However, ultimately this work is a starting point to a more advanced study and application of matrix-free optimization to electronic structure methods.
In future work, we plan to integrate localization and randomized linear algebra techniques to reduce computational prefactors and introduce optimization acceleration techniques such as the DIIS line-search method.\cite{Yu:2025:CILS}
As this is a preliminary study into the application of matrix-free decompositions to replace tensor-network contractions, we focus less on the reduced-scaling impacts of this work and more on the effects our method has on the CCSD residual equation and subsequent optimization accuracy.
In this next section we develop a new and reduced-scaling metric to terminate the matrix-free CP-PPL ALS minimization.

\subsubsection{\label{sec:cpdstop}Low cost CPD stopping condition}

The canonical stopping condition for the CPD is based on the change in the fit of approximation.
The fit of the approximated PPL tensor is
\begin{align}\label{eq:cpfit}
    \Delta_n = 1.0 - \frac{\| B - \overset{\approx}{B}{}_n\|}{\|B\|}
\end{align}
and, therefore, the canonical stopping condition is
\begin{align}\label{eq:canon}
    \Delta_n - \Delta_{n-1} < \epsilon
\end{align}
where $\epsilon$ is the chosen threshold value.
For the CP-PPL, computing the fit is prohibitively expensive (with a complexity of $\mathcal{O}(N^6)$).
We will break down the cost of the canonical stopping condition.

The numerator of \cref{eq:cpfit}, also defined as the residual, can be expanded as
\begin{align}
    \| B - \overset{\approx}{B}{}_n \| = \sqrt{B^2 - 2 * B \overset{\approx}{B}{}_n -\overset{\approx}{B}{}_n^2}.
\end{align}
The second and third terms under the square root can be computed with relatively low-cost;
the term $B\overset{\approx}{B}{}_n$ is partially-formed during the computation of the CPD gradient and formally has a complexity of $\mathcal{O}(N^5)$, as demonstrated previously.
While, the term $\overset{\approx}{B}{}_n^2$ can be computed by leveraging the CPD format
\begin{align}\label{eq:cp_norm2}
    \overset{\approx}{B}{}_n \overset{\approx}{B}{}_n =  \sum_{rr^\prime}[((\alpha_n)^T \alpha_n ) * ((\beta_n)^T\beta_n ) * ((\gamma_n)^T\gamma_n ) * ((\rho_n)^T) \rho_n ]^r_{r^\prime}.
\end{align}
and has a complexity of $\mathcal{O}(N^3)$.
Unfortunately, any attempt to construct $B$ or $B^2$ results in a $\mathcal{O}(N^6)$ computational complexity.

In principle, it would be possible to forgo a check completely and simply require a certain number of ALS iterations be completed.
However, it would be preferred to have some metric to effectively stop the optimization in as few steps as possible.
Therefore, in this work we introduce the new metric
\begin{align}\label{eq:N3_stopping}
    \eta = \frac{\| \overset{\approx}{B}{}_n - \overset{\approx}{B}{}_{n-1} \| }{\| \overset{\approx}{B}{}_{n-1} \|}
\end{align}
The numerator of $\eta$ can be expanded as 
\begin{align}\label{eq:new_conv}
    \eta_{\text{num}} = \sqrt{\overset{\approx}{B}{}_n^2 - 2 \overset{\approx}{B}{}_n\overset{\approx}{B}{}_{n-1} + \overset{\approx}{B}{}_{n-1}^2)} 
\end{align}
and all terms in this residual can be efficiently computed in $\mathcal{O}(N^3)$ time using the same mechanism found in \cref{eq:cp_norm2}.
The CPD is then terminated when $\eta < \epsilon$ for some chosen $\epsilon$ and an appropriate value of $\epsilon$ will be discussed in \cref{sec:results}.  
As with most optimization problems, the quality and duration of the CPD optimization can strongly depend on the initial guess.
Therefore, in the following section we will propose an improved initial guess strategy to minimize the required number of ALS iterations and therefore the cost of computing the CPD.

\subsubsection{\label{sec:cpdinit}An improved initial guess strategy}
In general, nonlinear optimizations require an unknown (potentially large) number of iterations and the number of iterations can depend strongly on the quality of an initial guess.
In this application we require the CPD optimization be computed every iteration of the CCSD optimization.
It is, therefore, in our best interest to find a high-quality initial guess strategy that minimizes the number of CPD optimization iterations.

In our previous work,\cite{VRG:pierce:2021:JCTC} we have found that a reasonable starting guess is a collection of random numbers chosen from a uniform distribution between $[-1, 1]$.
In this work, too, we see that this guess puts the CPD solver in a good position to converge to a relatively accurate optimization in as few as 13 iterations.
However, consider the following, if we choose a random guess for each CP-PPL optimization during a CCSD optimization, we are effectively treating each PPL tensor-network as an independent CPD problem.
We hypothesize that, for a single CCSD calculation, the optimized CP-PPL factor matrices during the $n$th CCSD iteration are close in optimization space to the CP-PPL factor matrices of the $m>n$ CCSD iteration. 
In this way, we are positing that there is some relationship between optimized CPD approximations of a given tensor-network, which are components of an iterative gradient optimization method, such as CCSD.
Therefore our proposed strategy is to find and cache a set of optimized CP-PPL factor matrices during the $n$th CCSD iteration to use as an initial guess for a later CP-PPL optimization.
A more complete description of our initial guess procedure will be presented and analyzed in \cref{sec:results}.
Next, we attempt to put our approach in perspective of previous studies that apply low-rank tensor decompositions to the CC family of methods.

\subsection{Literature Perspective}\label{sec:lit_rev}
In electronic structure methods, there exists only a small number of studies that utilize analytically constructed CPD tensors\cite{VRG:benedikt:2011:JCP,Benedikt:2013:JCP,Benedikt:2013:MP,Bohm:2016:JCP,Schmitz:2017:JCP,Khoromskaia:2015:PCCP,Madsen:2018:JCP} and thus, the utility of the analytical CPD in the field is still an open question.
In their previous work, Auer et al. replaced practically every high-order tensor with a CP approximation. 
although these authors formally achieve a reduced-scaling implementation, Auer concedes that these CP-based approaches are slowed significantly by the  spontaneous growth of the CP rank which results in the need for periodic re-optimization of CP approximated tensors.
Recently, Madsen et al.\cite{Madsen:2018:JCP} use a similar idea in the vibrational coupled cluster (VCC) method. 
While their work, again, requires the periodic re-optimization of CP factor matrices, these optimizations can be computed relatively efficiently and the authors also take advantage of the Tucker decomposition to further reduce the cost of their CPD optimization.\cite{Schmitz:2017:JCP}
Overall Madsen finds a significant benefit with the introduction of the CPD to the VCC method.
In this work, we explore a new application of the analytic CPD to directly approximate high-scaling tensor-networks without approximating components of the networks.
In a sense, this manuscript serves to extend Auer and Madsen's work by preserving the structure of a tensor-network using a fast-update scheme that simultaneously contracts the tensor-network and optimizes a low-rank decomposition.
It should be noted that in this preliminary study, we do not preserve the decomposed tensor-network format and choose instead to reconstruct the result of our decomposition.

The core idea of this manuscript parallels the idea presented by Schutski et al.\cite{Schutski:2017:JCP}
In their work, the authors propose a means to optimize a THC representation of the two-particle excitation tensor while simultaneously minimizing the CCSD equations, \cref{eq:ccsd}.
In a similar manner to the proposed method in this work, the authors first create the loss function
\begin{align}\label{eq:GusTHCT2}
    f(\hat{T_2}) = \frac 1 2 \|T_2 - \tilde{T_2}\|^2
\end{align}
where $\tilde{T_2}$ is a THC representation of the two-particle excitation tensor.
Schutski then replaces the exact $T_2$ tensor with a tensor-network representation obtained from the canonical update of the tensor, i.e.
\begin{align}
    t^{ab}_{ij} = \frac{\langle\Phi^{ab}_{ij} | \bar{H} | \Phi_0\rangle}{\epsilon_a + \epsilon_b - \epsilon_i -\epsilon_j}.
\end{align}
By replacing the exact $T_2$ tensor with this tensor-network approximation the authors develop a matrix-free optimization of $\hat{T}_2$ and, like we demonstrate in this work, are able to reduce the complexity of tensor-network contractions in the residual equation.
While this matrix-free optimization is functionally equivalent, Schutski's optimization process is made complicated by the sheer number of tensor-networks which must be evaluated to form the least-square solutions to \cref{eq:GusTHCT2} and the fact that the CCSD residual tensor, $R^{ab}_{ij} = \langle\Phi^{ab}_{ij} | \bar{H} | \Phi_0\rangle$, is itself a function of $T_2$.
In that work, the authors choose to replace the $T_2$ in each iteration with the optimized approximation $\hat{T}_2$ from the previous CCSD iteration.
In this work, we effectively reduce the scope of Schutski's work to understand the role of error propagation in approximated tensor-networks and to demonstrate that individual components of the CCSD residual equation are low-rank tensors.
In principle, we could reformulate the optimization of $T_2$ presented by Schutski and replace the THC approximation with a CPD; however, practically, this would also lead to an equally large and complicated set of tensor-network intermediates. 
Alternatively, we are currently investigating efficient means to combine these studies to accurately accelerate the CC family of methods.

The work by Schutski motivated the development of the rank-reduced coupled cluster method.\cite{Parrish:2019:JCP,Hohenstein:2019:JCP,Hohenstein:2021:JCP,Hohenstein:2022:JCP,Lesiuk:2022:JCP}
This approximation also represents the excitation amplitude tensors in a low-rank THC decomposition. 
However, unlike Schutski's tensor-structured decomposition, the rank-reduced approximation only determines updates to the THC core tensor. 
By doing so, this rank-reduced approximation effectively reduce the number of parameters in the gradient optimization of CCSD from $N^4$ to $N^2$ and in coupled cluster with single, double and triple excitations (CCSDT) from $N^6$ to $N^3$ using a projection-based approach.
This idea creates relatively compact CC algorithms compared to Schutski's tensor-structured implementation.
Next, we describe the computational experiments used to validate this work.

\section{\label{sec:compdetails} Computational details}
All CCSD calculations have been computed using a developmental version of the massively parallel quantum chemistry (MPQC) software package\cite{VRG:calvin:2020:CR,Peng:2020:JCP} 
and computations were run on the Flatiron Institute Scientific Computing Core's (SCC) Rusty cluster using the standard Rome nodes that are equipped with 2 AMD EPYC 7742 processors that each have 64 CPU cores and 512 GB of memory. 
Each CC calculation was run on a 16 core partition of a single Rome CPU with each partition containing 128 GB of memory. 
The CP approximation of the order-4 PPL tensor-network was computed using a standard alternating least squares (ALS) method,\cite{VRG:kroonenberg:1980:P,VRG:beylkin:2002:PNAS} as discussed previously.
Using the initialization schemes discussed in \cref{sec:cpdinit}, we see rapid convergence to high accuracy approximation and the full extent of these optimizations will be discussed in \cref{sec:results}.
All CPD optimizations that utilize the canonical CP stopping condition, \cref{eq:cpfit}, use a convergence criteria of $\epsilon = 10^{-3}$, which was found to be sufficiently accurate in our previous published results.\cite{VRG:pierce:2021:JCTC,VRG:pierce:2023:JCTC}
The focus of our study is small water clusters with between 1 to 6 molecules in the TIP4P optimized geometry\cite{jorgensen:1983:JCP,Wales:1998:CPL} from the website 
\begin{verbatim}
    http://www-wales.ch.cam.ac.uk/~wales/CCD/TIP4P-water.html
\end{verbatim}
and small alkane chains with between 1 and 6 carbon atoms.
We analyze the different initial-guess and convergence strategies and wall-time performance of the CPD using the water clusters in the cc-pVDZ-F12 (DZ-F12) orbital basis set (OBS), accompanied by the corresponding aug-cc-pVDZ-RI (aVDZ-RI) DF basis set (DFBS).\cite{VRG:peterson:2008:JCP,VRG:weigend:2002:JCP}
We choose this basis for its intermediate size between the double and triple zeta basis.
We also analyze the approach using the cc-pVTZ (TZ)\cite{Dunning1989,Kendall1992} OBS accompanied with the corresponding cc-pVTZ-RI (TZ-RI)\cite{Weigend2002a} DFBS to determine the scaling of the CP rank with system size.
In the following section, we present our findings from these experiments.

\section{\label{sec:results} Results}

In this section, we demonstrate that it is possible to accurately replace the exact computation of the CCSD PPL tensor-network with a low-rank, matrix-free tensor factorization. 
Further, we establish an understanding of the following CP-PPL parameters: initial guess strategy, CP rank and CPD stopping condition, and we consider the computational limitations of this approach.

\begin{figure}[t]
    \begin{subfigure}{0.5\textwidth}
        \includegraphics[width=\columnwidth]{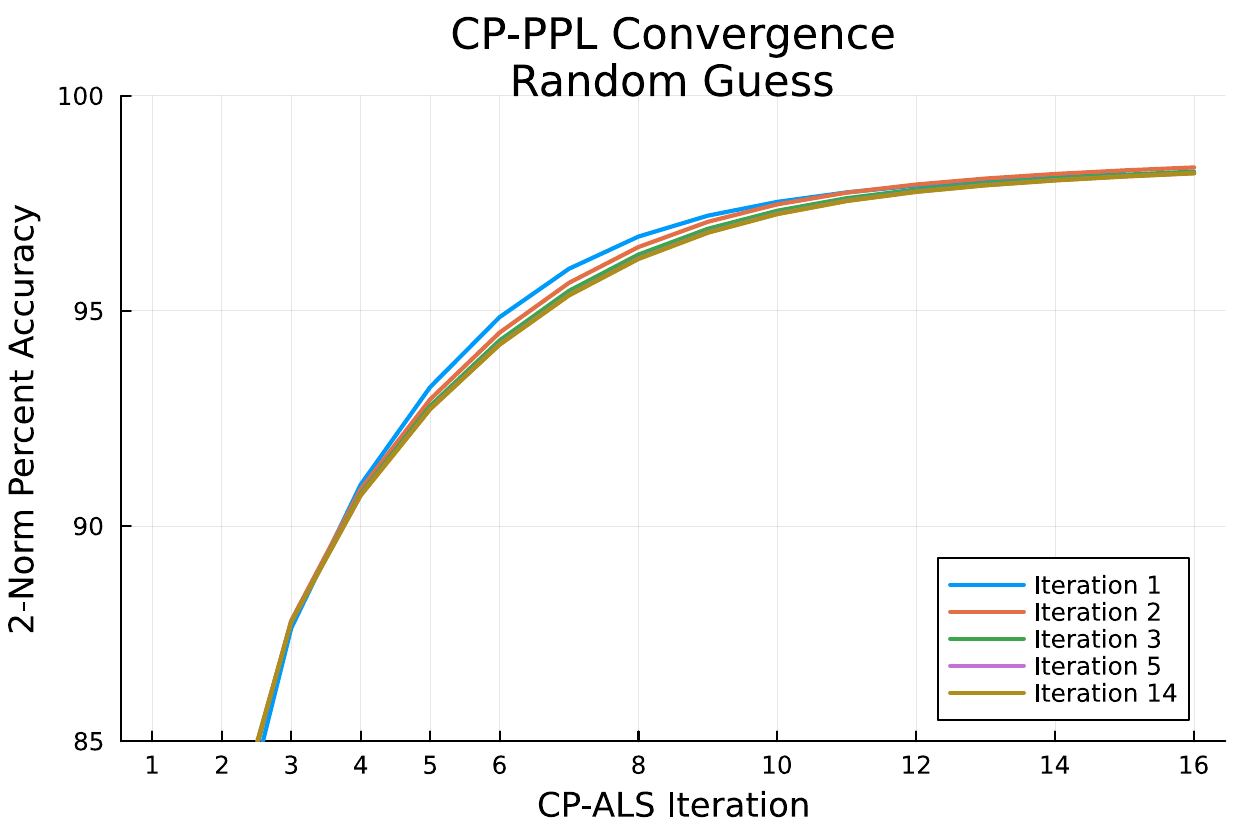}
        \caption{}
        \label{fig:rand_guess_als_conv}
    \end{subfigure}\hfill
    \begin{subfigure}{0.49\textwidth}
        \includegraphics[width=\linewidth]{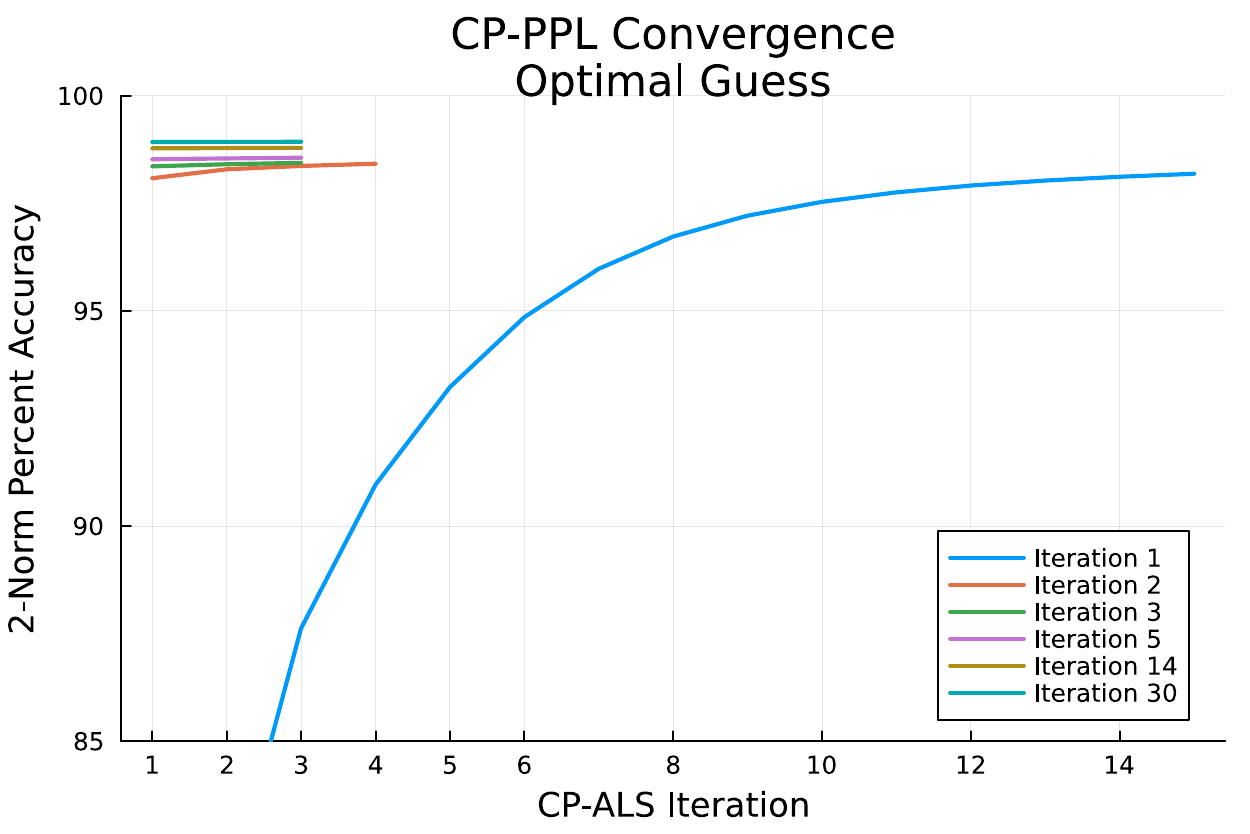}
        \caption{}
        \label{fig:opt_guess_als_conv}
    \end{subfigure}\hfill
    \begin{subfigure}{0.49\textwidth}
        \includegraphics[width=\linewidth]{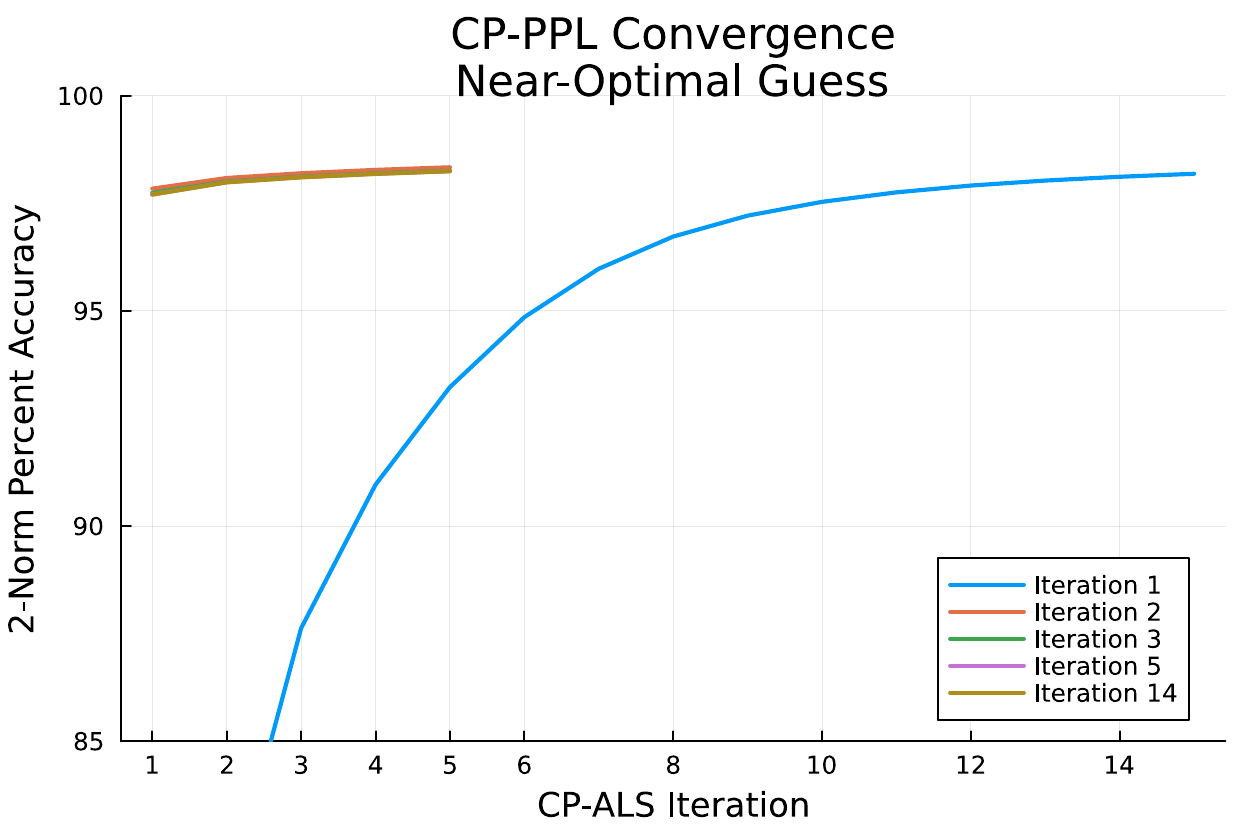}
        \caption{}
        \label{fig:no_guess_als_conv}
    \end{subfigure}\hfill
\caption{Convergence of the CP-PPL ALS optimization using the (a) "Random guess", (b) "Optimal guess" and (c) "Near-Optimal guess" for a 2 water molecule cluster in DZ-F12/aVDZ-RI with a fixed CP rank of $3.5X$ during different CCSD iterations.}
\label{fig:als_conv}
\end{figure}

\subsection{Role of the Initial Guess}
First, we investigate the role of the CP-PPL initial guess and study its effects on the CCSD optimization.
In this work we use three different initial guess strategies:
\begin{itemize}
    \item with "Random guess" we initialize each CPD with a set of randomly initialized factor matrices;
    \item with "Optimal guess" we initialized every CPD with the output of the previous CPD optimization; and
    \item with "Near-Optimal guess" we initialize the current CPD with a cached set of factor matrices from a previous optimization. The caching procedure is as follows: if the current CPD optimization requires $n>5$ iterations, then the factor matrices from the $(n-5)$th iteration are cached. Else, the current optimized factor matrices are not cached for later use and a previously cached set is used in the subsequent optimization.
\end{itemize}

\begin{figure}
    \begin{subfigure}{0.49\textwidth}
        \includegraphics[width=\linewidth]{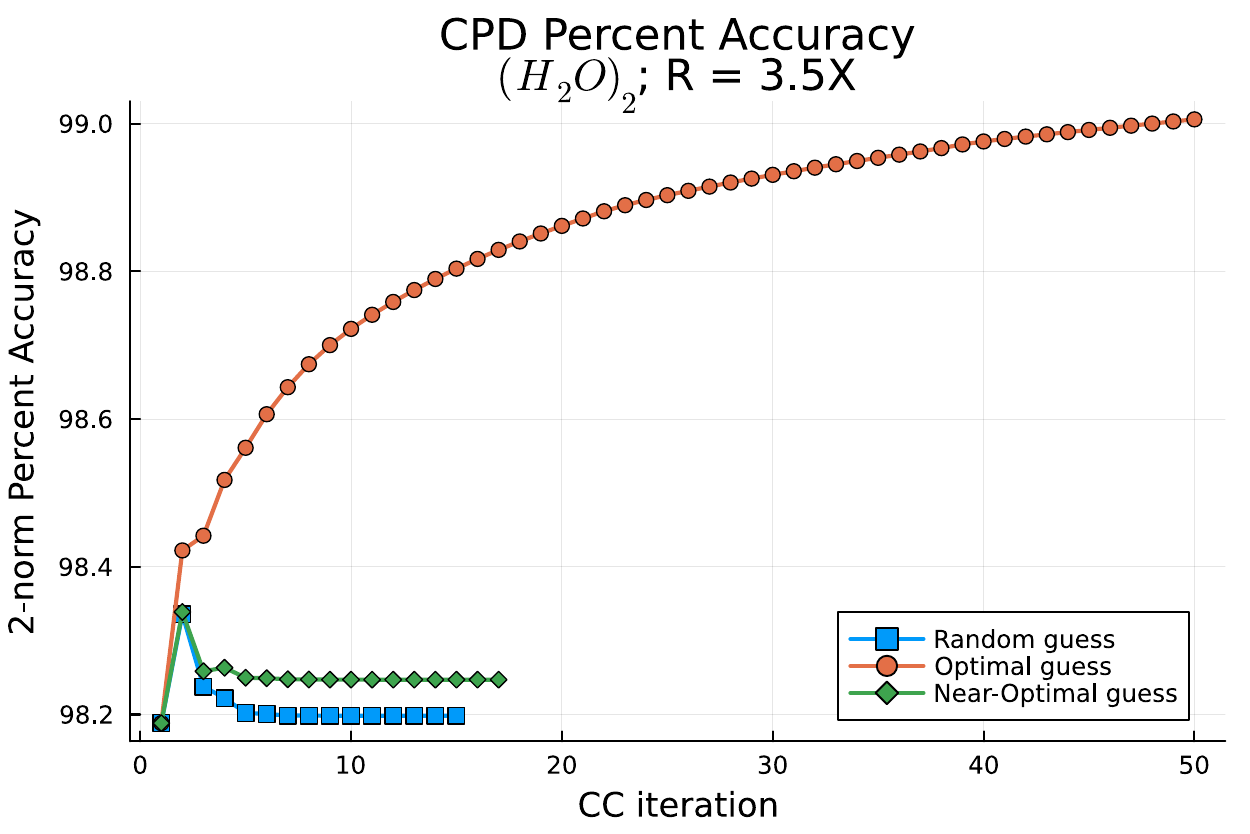}
        \caption{}
        \label{fig:CPD_CC_Opt}
    \end{subfigure}\hfill
    \begin{subfigure}{0.5\textwidth}
        \includegraphics[width=\columnwidth]{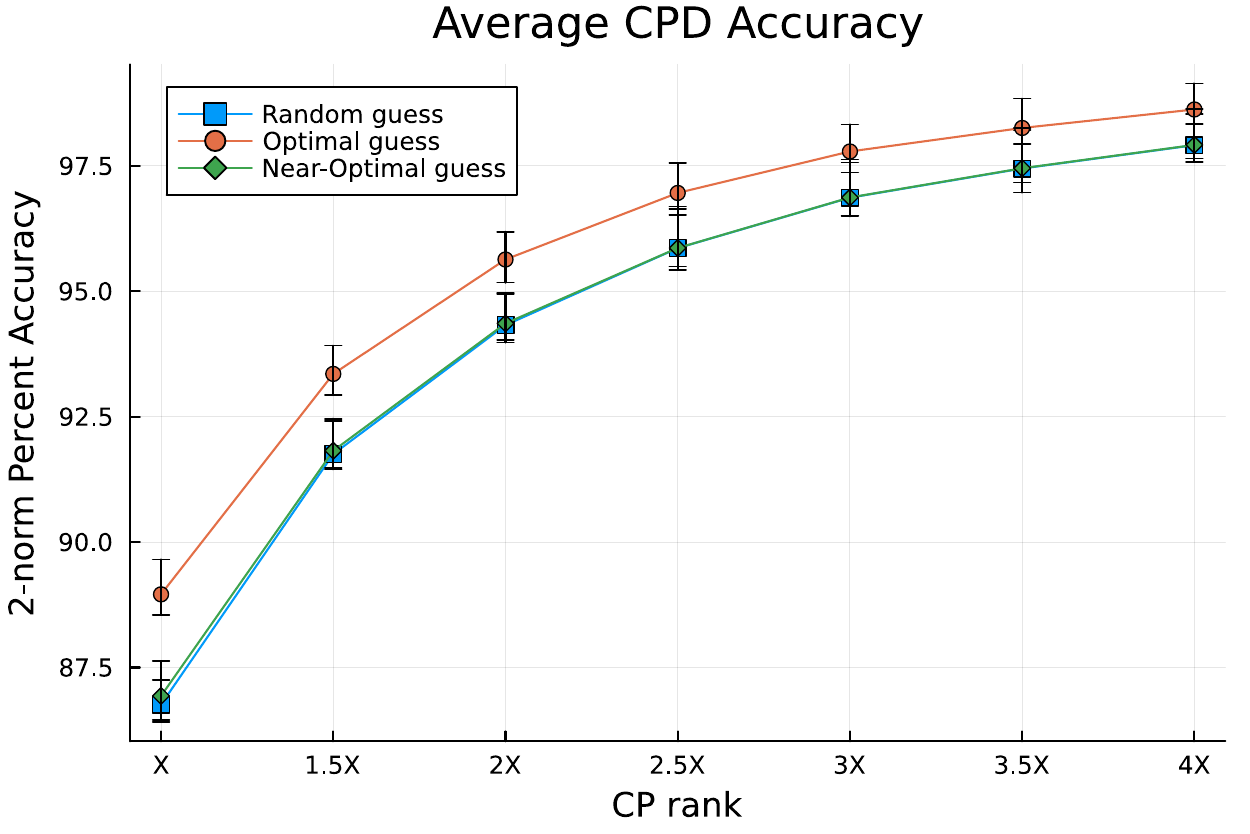}
        \caption{}
        \label{fig:AvgCPDAcc}
    \end{subfigure}\hfill
\caption{(a) CP-PPL percent accuracy across a CCSD optimization for a two water molecule cluster in DZ-F12/aVDZ-RI at fixed CP rank of $3.5X$ using the three CPD initialization schemes. (b) CP-PPL percent accuracy for clusters with between 1 and 5 water molecules in DZ-F12/aVDZ-RI using three CPD initialization schemes. Error bars represent maximum and minimum percent accuracy across water clusters.}
\end{figure}

For all of these schemes, the first CPD optimization of a network always uses the "Random guess."
For the "Near-Optimal guess" scheme the value $5$ was chosen to be close enough to the end of the optimization for future optimizations to converge quickly and far enough away to consistently optimize the CP-PPL to nearly the same 2-norm accuracy, as one can see in \cref{fig:no_guess_als_conv,fig:CPD_CC_Opt}.

In \cref{fig:als_conv}, we illustrate the convergence of the CP-PPL ALS optimization at different iterations of CCSD using the three different initial guess strategies. 
As mentioned previously, we define convergence as when the change in fit, \cref{eq:canon}, is less than $10^{-3}$.
In \cref{fig:als_conv} we plot the 2-norm percent accuracy (which is computed as 100 times the 
CP fit, \cref{eq:cpfit}) for each CP-ALS iteration in lieu of the change in CP fit. 
Therefore, in this figure, convergence correlates with the slope between consecutive points.
As discussed, all three methods use the "Random guess" for the first iteration and these optimizations converge consistently in about 13 iterations. 
With \cref{fig:rand_guess_als_conv}, one can that this "Random guess" consistently converges in roughly 13 iterations for each iteration of the CCSD algorithm and the accuracy of the CP-PPL is consistent across iterations.
With \cref{fig:no_guess_als_conv}, one can see that the "Near-Optimal guess" significantly reduces the number of ALS iterations from 13, using the "Random guess," to 5.
And, like the "Random guess," the accuracy of the CP-PPL using the "Near-Optimal guess" is consistent every iteration of CCSD, unlike the "Optimal guess" strategy.

With \cref{fig:opt_guess_als_conv} one can see that the "Optimal guess" significantly reduces the number of ALS iterations to 4 or fewer.
Interestingly, with the "Optimal guess" the accuracy of the CP-PPL improves every CC iteration. 
A possible explanation for this phenomenon is that from iteration 3 and beyond, the PPL tensor does not dramatically change.
Therefore, the "Optimal guess" strategy continues to improve the CP-PPL during each CCSD iteration, converging the approximation far beyond the prescribed ALS convergence criteria.
If this is the case, then this result demonstrates that there is room for improvement in the fixed-rank CP-PPL optimization.
One route to improve the convergence of the ALS solver is using line-search methods such as the Direct Inversion of the Iterative Subspace method;\cite{Yu:2025:CILS} alternatively we could employ more efficient optimization strategies like the accelerated gradient descent technique.\cite{Acar:2011:JC,Phan:2013:SJMAP,Tichavsky:2013:ICASSP,VRG:benedikt:2011:JCP,Espig:2012:NM}

We illustrate the 2-norm percent accuracy of the CP-PPL across a CCSD optimization of a 2 water molecule cluster using the three initial guess strategies with a fixed CP rank of $3.5X$ in \cref{fig:CPD_CC_Opt}.
The results in \cref{fig:CPD_CC_Opt} corroborate the results from \cref{fig:als_conv} in that the 2-norm accuracy of the CP-PPL is consistent across a CCSD optimization using the "Random guess" and "Near-Optimal guess" strategies but improves consistently using the "Optimal guess" strategy.
In \cref{fig:AvgCPDAcc}, we plot the averaged 2-norm based percent accuracy of the CP-PPL across water clusters with between 1 and 5 molecules and averaged across all CCSD iterations.
To compile the data in \cref{fig:AvgCPDAcc} we first average the CPD accuracy over the number of CCSD iterations for each molecule.
Then we average the averaged CPD accuracies over the set of molecules tested.
This figure demonstrates that, as expected, increasing the CP rank directly improves the accuracy of the approximation of $B$. 
Similar to the previous results, \cref{fig:AvgCPDAcc} shows that the "Random" and "Near-Optimal" guesses produce roughly the same CP-PPL approximation conversely, the "Optimal guess" initial guess scheme produces a better CPD optimization on average than the other two initial guess schemes.
Also in \cref{fig:AvgCPDAcc}, we see that the averaged CP-PPL accuracy is rather consistent with molecular system size. 

\begin{figure}
    \begin{subfigure}{0.49\textwidth}
        \includegraphics[width=\linewidth]{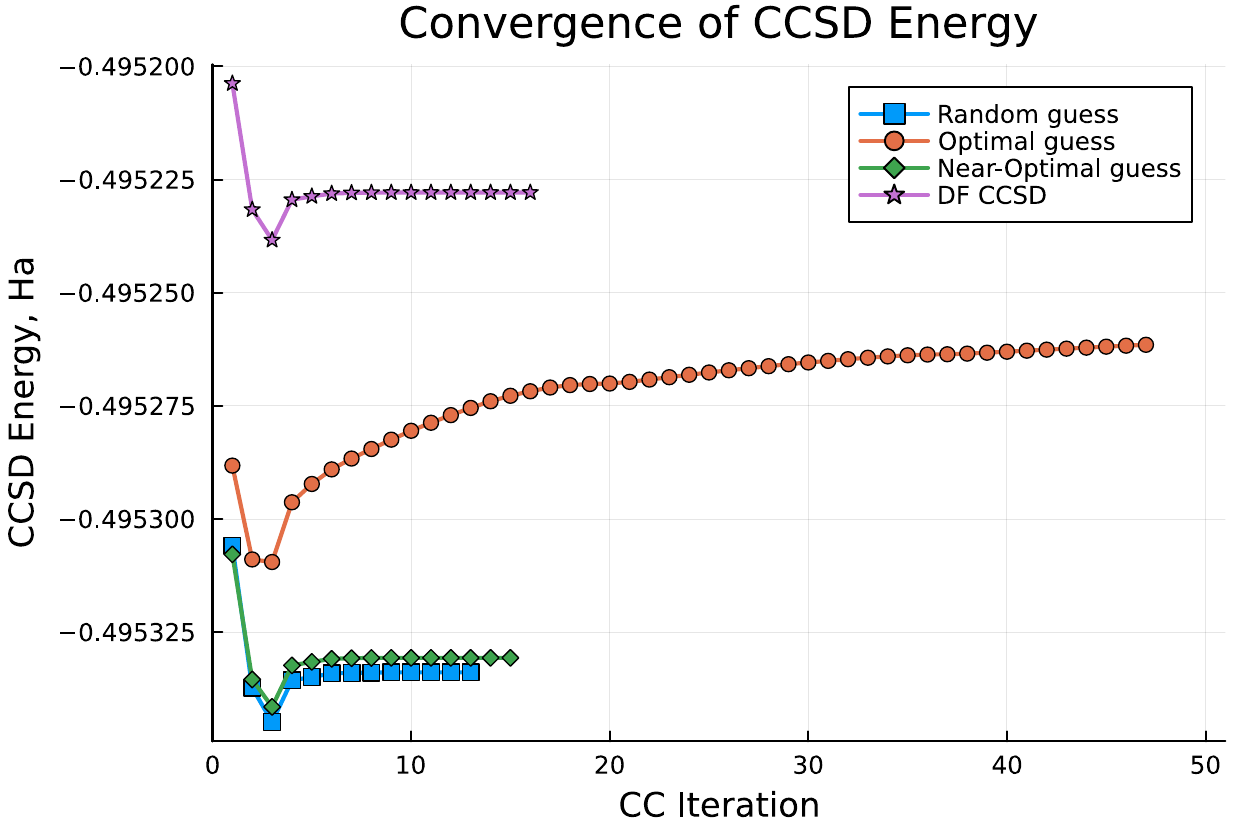}
        \caption{}
        \label{fig:3a}
    \end{subfigure}\hfill
    \begin{subfigure}{0.5\textwidth}
        \includegraphics[width=\columnwidth]{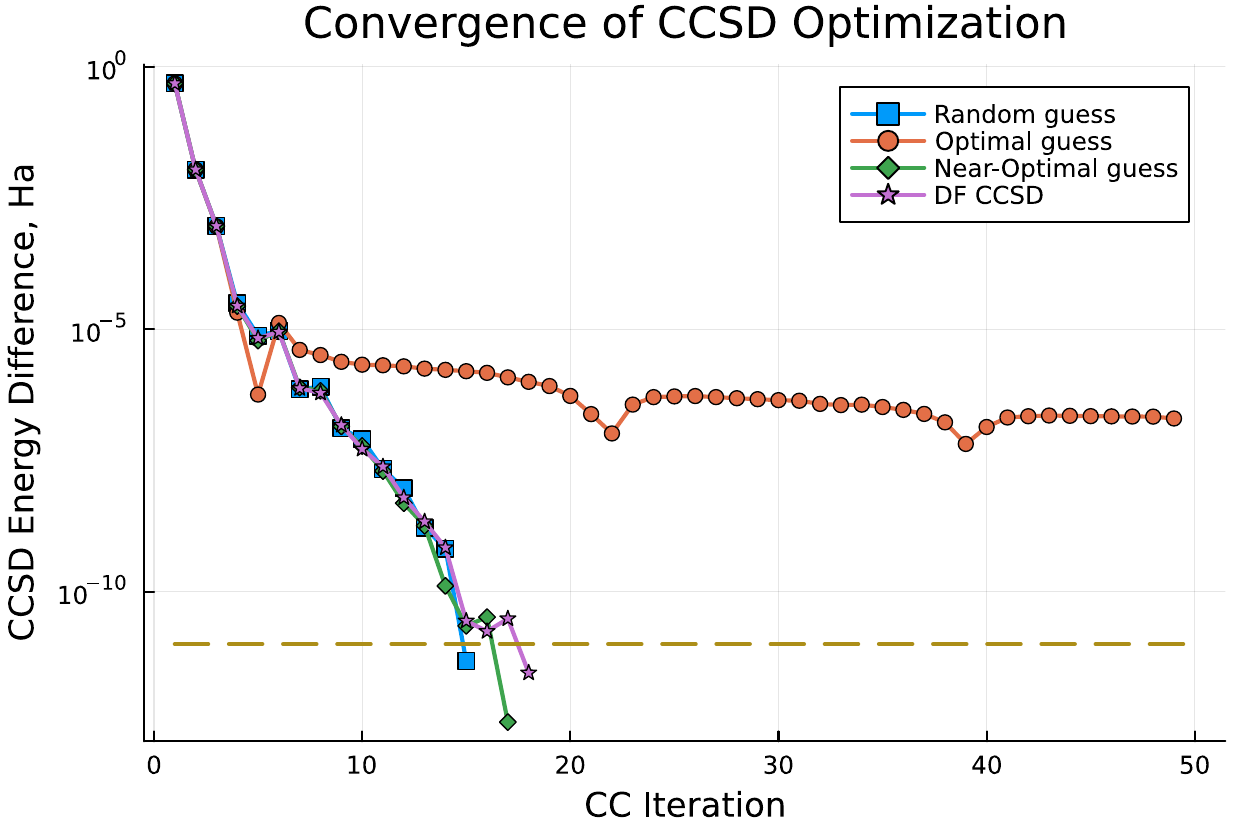}
        \caption{}
        \label{fig:3b}
    \end{subfigure}\hfill
\caption{Comparing the CCSD energy convergence using (a) absolute energy and (b) per iteration energy differences for a two water molecule cluster in DZ-F12/aVDZ-RI and CP-PPL calculations using a fixed CP rank of $3.5X$. The dashed line in (b) represents the CCSD energy convergence criteria.}
\label{fig:CC_conv}
\end{figure}

One should notice that in \cref{fig:CPD_CC_Opt} the 
"Optimal guess" strategy is computed for 50 CC iterations, whereas the other strategies require fewer CC iterations.
We investigate this further in \cref{fig:CC_conv}.
In our calculations, CCSD convergence is determined when the change in CCSD energy falls below $10^{-11}$ Ha.
In this figure, one can see that the "Random guess" and "Near-Optimal guess" strategies allow CCSD to converge quickly to this threshold, requiring roughly the same number of CCSD iterations as the canonical calculation, here denoted "DF CCSD."
One can see that with the "Optimal guess" scheme, because the CP-PPL improves consistently throughout the course of the CCSD optimization, the CCSD energy also improves consistently throughout the optimization.
This behavior is not exactly desirable as it causes the CC optimization to "swamp" and never reach the target energy convergence of $10^{-11}$ Ha.
In no single CP-PPL based CCSD optimization computed with the "Optimal guess" scheme does the CCSD loop converge to the requested energy difference cutoff in 50 iterations.
Perhaps this suggests an issue with the robustness of the standard Jacobi-DIIS CCSD solver and that such a solver only performs well with consistently approximated terms in its gradient expression. 
Alternatively, this could be fixed using a tighter CP-PPL ALS convergence criteria or by stopping the CP-PPL optimization after a number of CCSD iterations.
In future work, we hope to garner a better understanding of the interplay between the CPD optimization/approximation of the PPL ladder diagram and the CC optimization landscape to resolve this swamping issue.
Next, we analyze the accuracy of the CP-PPL approximated CCSD method using these three initialization strategies.

\begin{figure}[t]
    \begin{subfigure}{0.49\textwidth}
        \includegraphics[width=\linewidth]{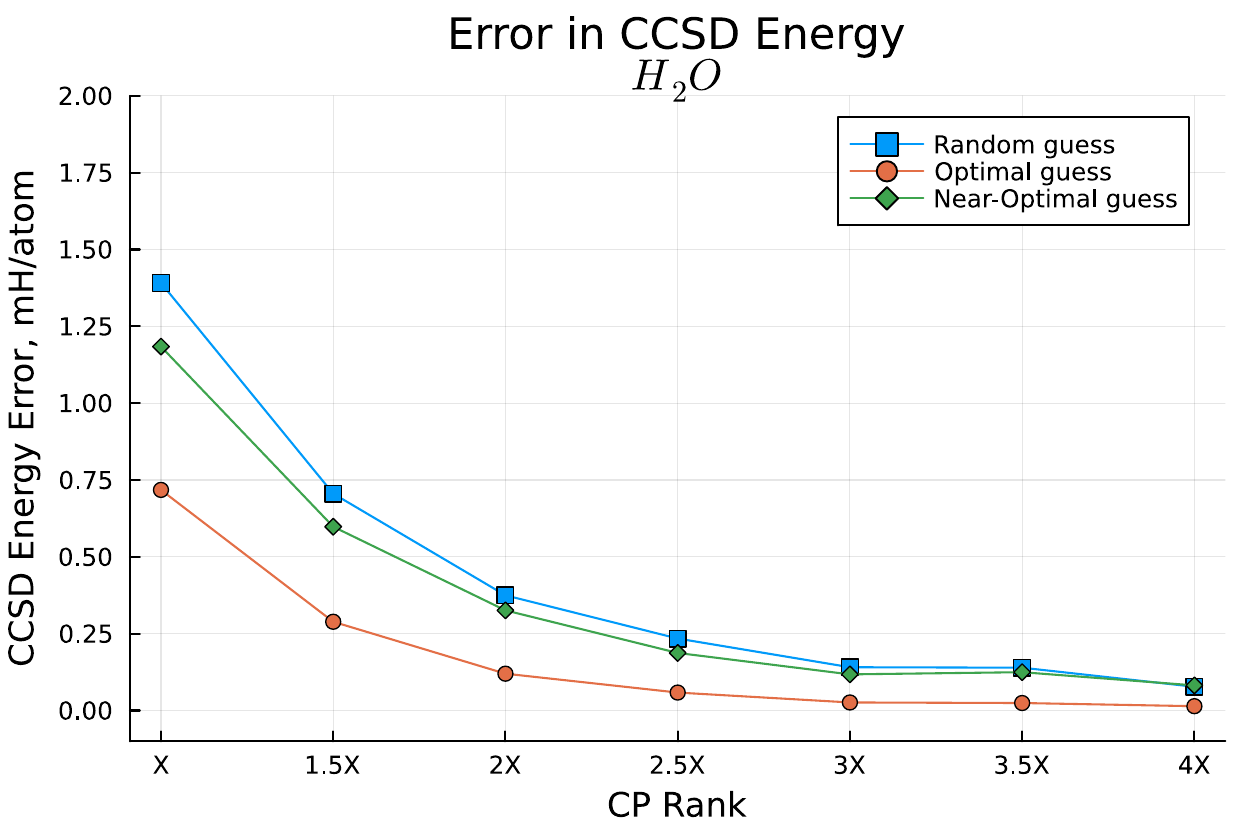}
        \caption{}
        \label{fig:1wat_en_conv}
    \end{subfigure}\hfill
    \begin{subfigure}{0.49\textwidth}
        \includegraphics[width=\linewidth]{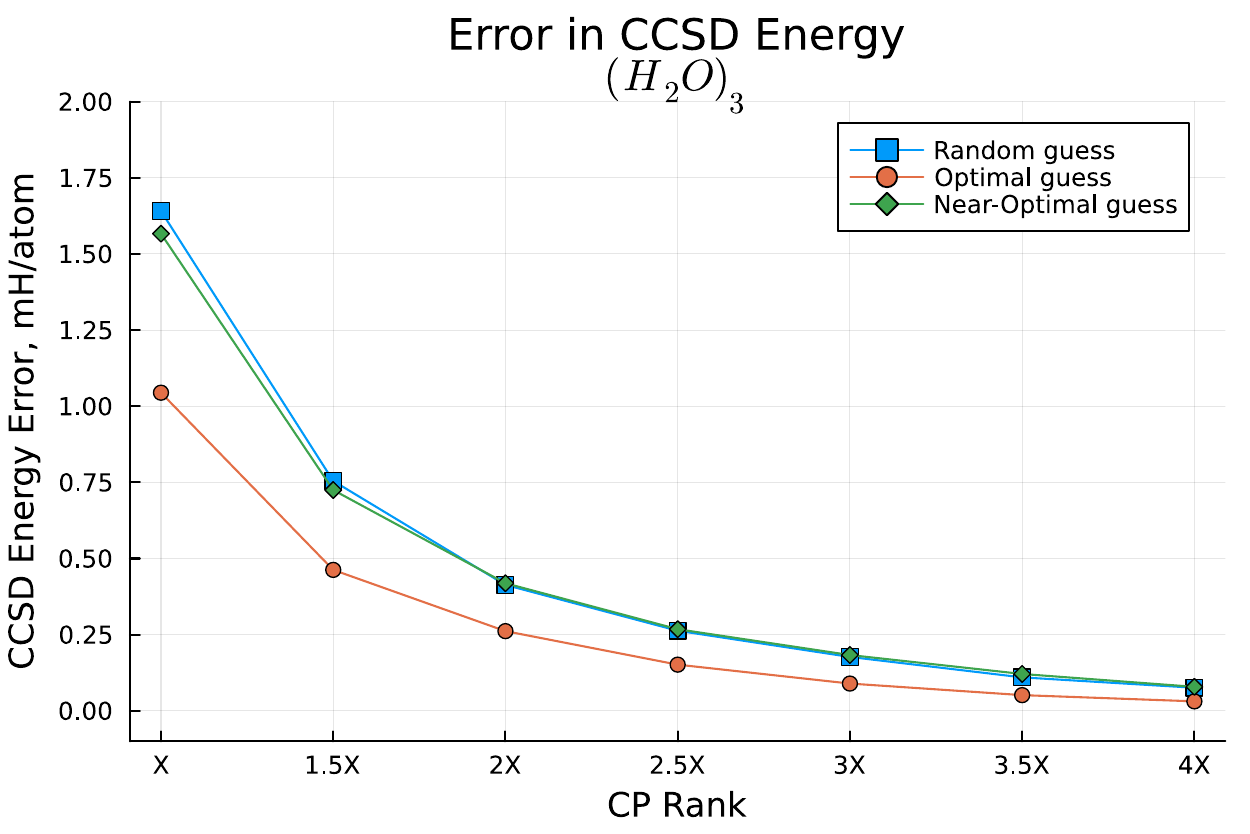}
        \caption{}
        \label{fig:3wat_en_conv}
    \end{subfigure}\hfill
    \begin{subfigure}{0.49\textwidth}
        \includegraphics[width=\linewidth]{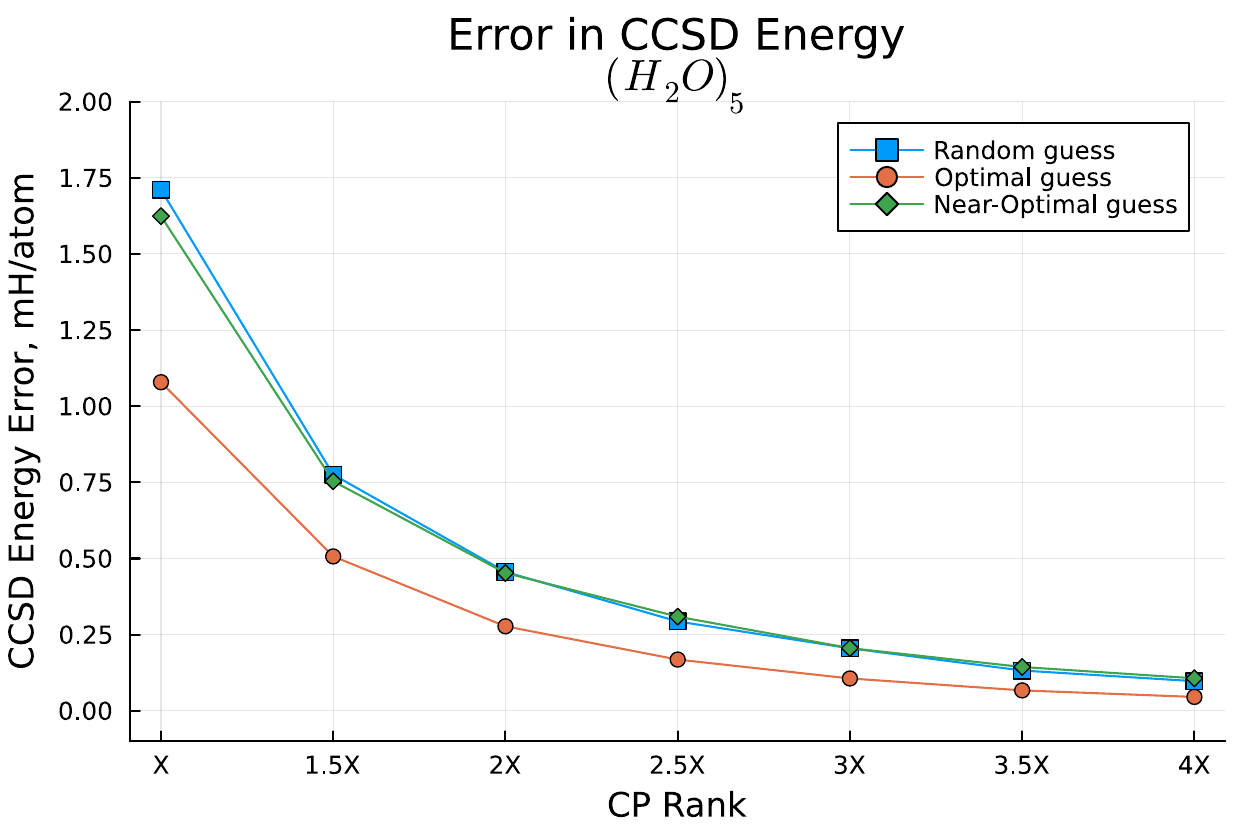}
        \caption{}
        \label{fig:5wat_en_conv}
    \end{subfigure}\hfill
    \caption{Absolute CCSD energy error per non-hydrogenic atom for clusters of (a) 1, (b) 3 and (c) 5 water molecules using the DZ-F12/aVDZ-RI basis versus CP rank using the various initial guess schemes. The "Random guess" and "Near-Optimal guess" schemes were converged to a CCSD energy difference of $10^{-11}$ Ha, however all "Optimal guess" computations did not converge in 50 CCSD iterations.}
\label{fig:AvgPerElec}
\end{figure}

\subsection{Analyzing the accuracy of the CPD-PPL}
Here we show that sufficiently accurate and systematically improvable energy values can be computed by replacing the PPL tensor-network contraction with the CP-PPL approximation.
First, \cref{fig:AvgPerElec} shows the error in the CP-PPL approximated CCSD energy compared to the canonical DF approximated CCSD energy for various water molecule clusters in the DZ-F12/aVDZ-RI basis using the three different initial guess strategies described in the previous section.
As the CP-PPL approximation improves, all initial guess strategies find increasingly accurate energy values.
The "Random guess" and "Near-Optimal guess" approaches both find nearly identical energy values while the "Optimal guess" scheme finds significantly more accurate CCSD energy values for every molecule and at every rank.
Also, there does not seem to be a dramatic difference in CCSD energy error for these different water molecule clusters.
\begin{figure}
    \begin{subfigure}{0.5\textwidth}
        \includegraphics[width=\columnwidth]{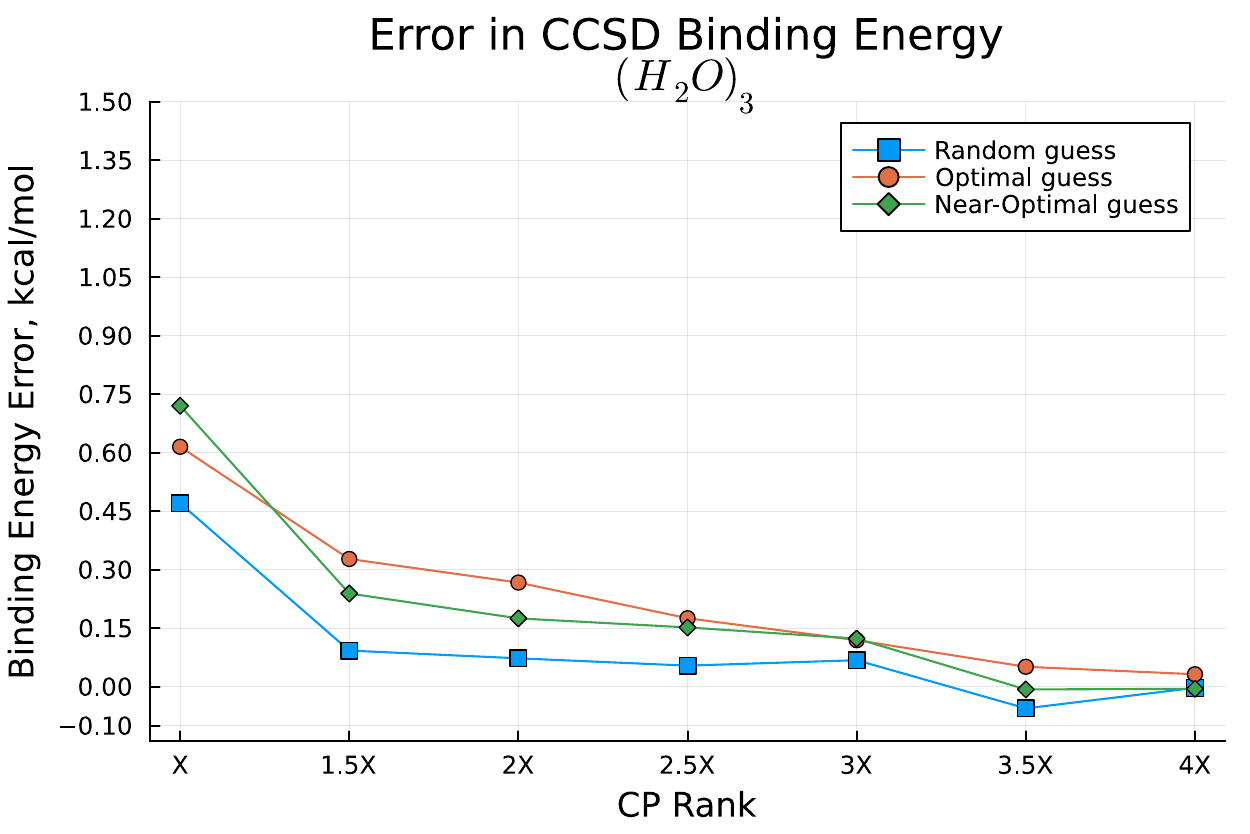}
        \caption{}
        \label{fig:signBindEner}
    \end{subfigure}\hfill
    \begin{subfigure}{0.49\textwidth}
        \includegraphics[width=\linewidth]{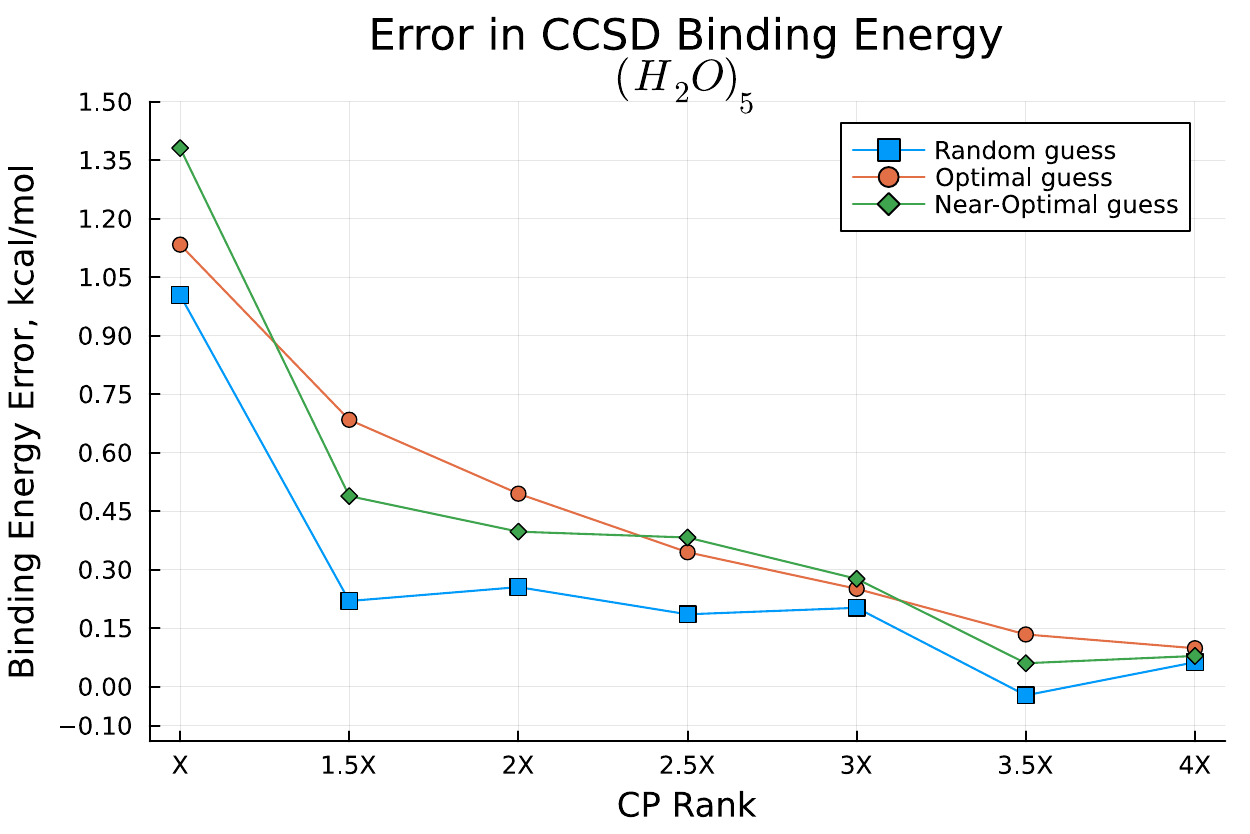}
        \caption{}
        \label{fig:absBindEner}
    \end{subfigure}\hfill
\caption{Signed binding energy error in kcal/mol versus CP rank for clusters of (a) 3 and (b) 5 water molecules in the DZ-F12/aVDZ-RI basis.} \label{fig:bindingE}
\end{figure}

\cref{fig:bindingE} shows that, although the "Optimal guess" finds more accurate CCSD energies, this does not categorically translate to more accurate energy differences.
In fact, in this example, the "Optimal guess" scheme always finds higher errors in the binding energy of water when compared to the "Random guess" scheme.
Most likely, the lower accuracy binding energies are related to the fact that the "Optimal guess" does not converge the CCSD energies beyond $10^{-6}$ Ha.
However, the other approaches may more systematically approximate the CCSD energy, resulting in a fortuitous cancellation of error.
From \cref{fig:bindingE}, one can see that only a modest CP rank of $1.5X$ is required to compute the binding energy with an accuracy beyond chemical accuracy, 1 kcal/mol.
But, we recommend a rank of $3.5X$ or greater to find an accuracy beyond 0.1 kcal/mol.
In the Supporting Information, readers can find additional results comparing the accuracy of the CP-PPL to the component-wise approximation of the PPL tensor-network via the order-4 CPD of the TEI tensor.
In the next section, we introduce our reduced-scaling convergence CPD criteria.
\begin{figure}[!b]
    \begin{subfigure}{0.5\textwidth}
        \includegraphics[width=\columnwidth]{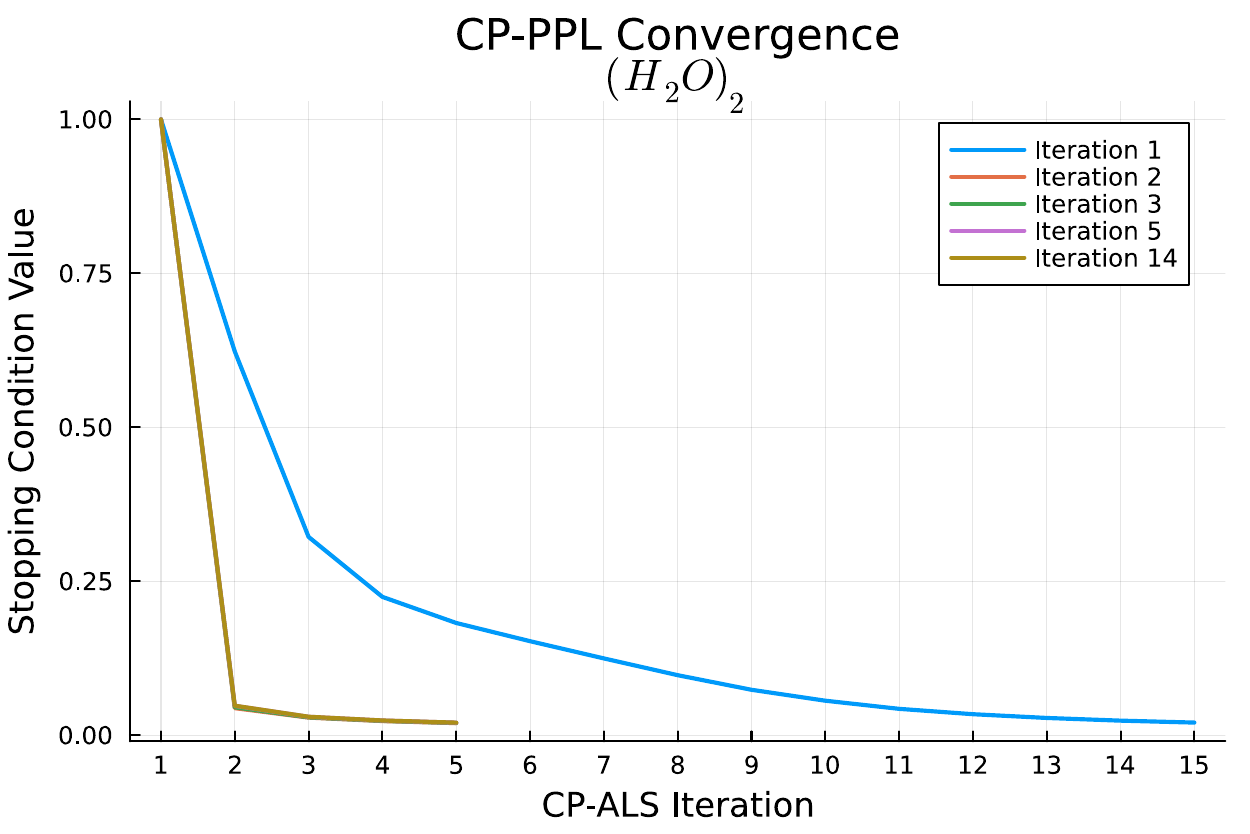}
        \caption{}
        \label{fig:cheap_conv_2w}
    \end{subfigure}\hfill
    \begin{subfigure}{0.49\textwidth}
        \includegraphics[width=\linewidth]{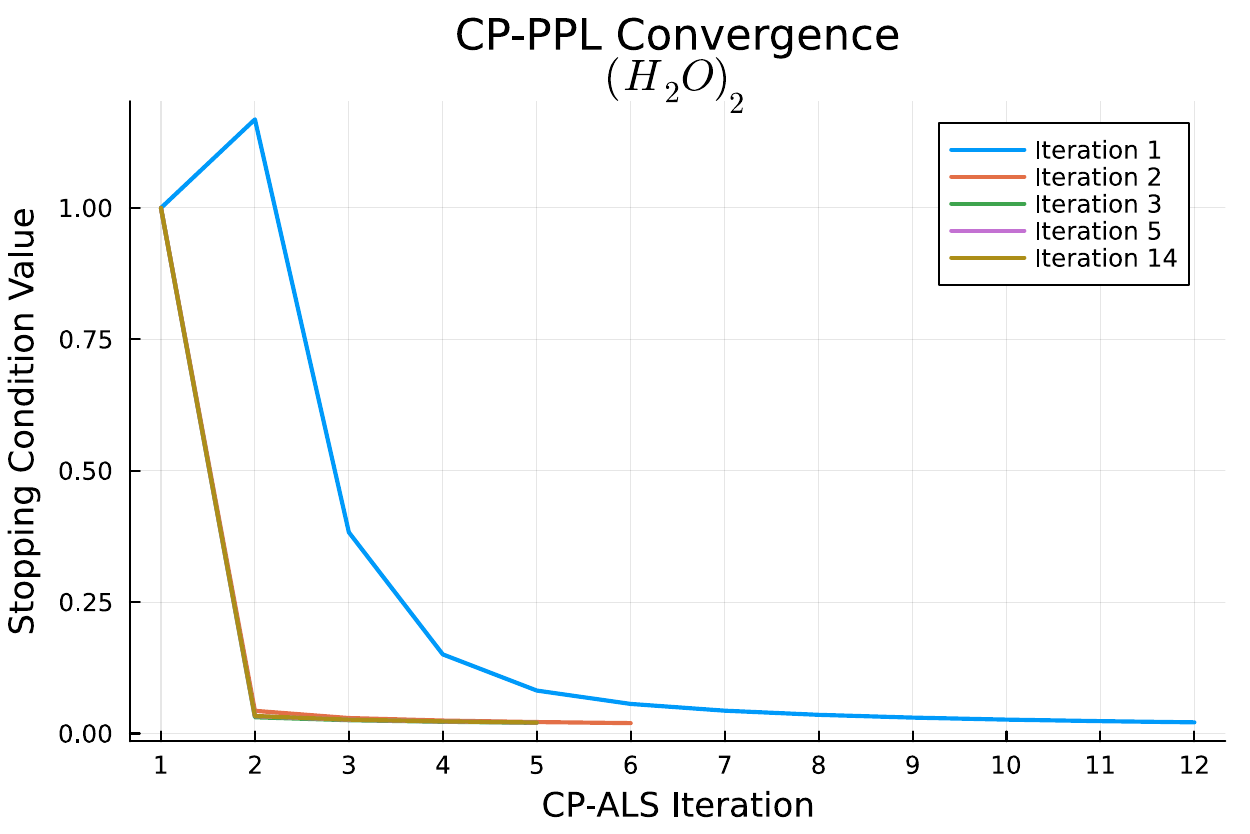}
        \caption{}
        \label{fig:cheap_conv_5w}
    \end{subfigure}\hfill
\caption{CP-PPL convergence via the evaluation of the reduced-scaling $N^3$ stopping condition for clusters of (a) 2 and (b) 5 water molecules in the DZ-F12/aVDZ-RI basis} \label{fig:cheap_als_conv}
\end{figure}

\subsection{Analyzing the Reduced-Scaling ALS Convergence Criteria }

Here we introduce our reduced-scaling CPD convergence criteria.
Thus far, our results have used the canonical stopping condition, \cref{eq:cpfit}.
As discussed previously, this stopping condition depends on the norm of the PPL tensor, which has an irreducible $\mathcal{O}(N^6)$ scaling.
In the figures to follow, we use the label "$N^3$ Stopping Condition" to denote the results using this new, reduced-scaling stopping condition from \cref{eq:N3_stopping}.
It should be noted that all calculations that use the reduced-scaling stopping condition also use the "Near-Optimal guess" scheme.
In general, the choice of stopping condition threshold, $\epsilon$, is an arbitrary decision and was chosen to replicate the balance between CP-PPL accuracy and number of least squares iterations found using the canonical stopping condition.
Here we do not perform a deep evaluation of the new stopping condition and choose the stopping condition of $\epsilon = 0.02418$ based on a simple heuristic: we choose a value of $\epsilon$ such that the water dimer molecule with a CP rank of $2X$ completes its CP-PPL optimization in 16 ALS iterations using the "Random Guess" initialization scheme.
\begin{figure}[!b]
    \begin{subfigure}{0.5\textwidth}
        \includegraphics[width=\columnwidth]{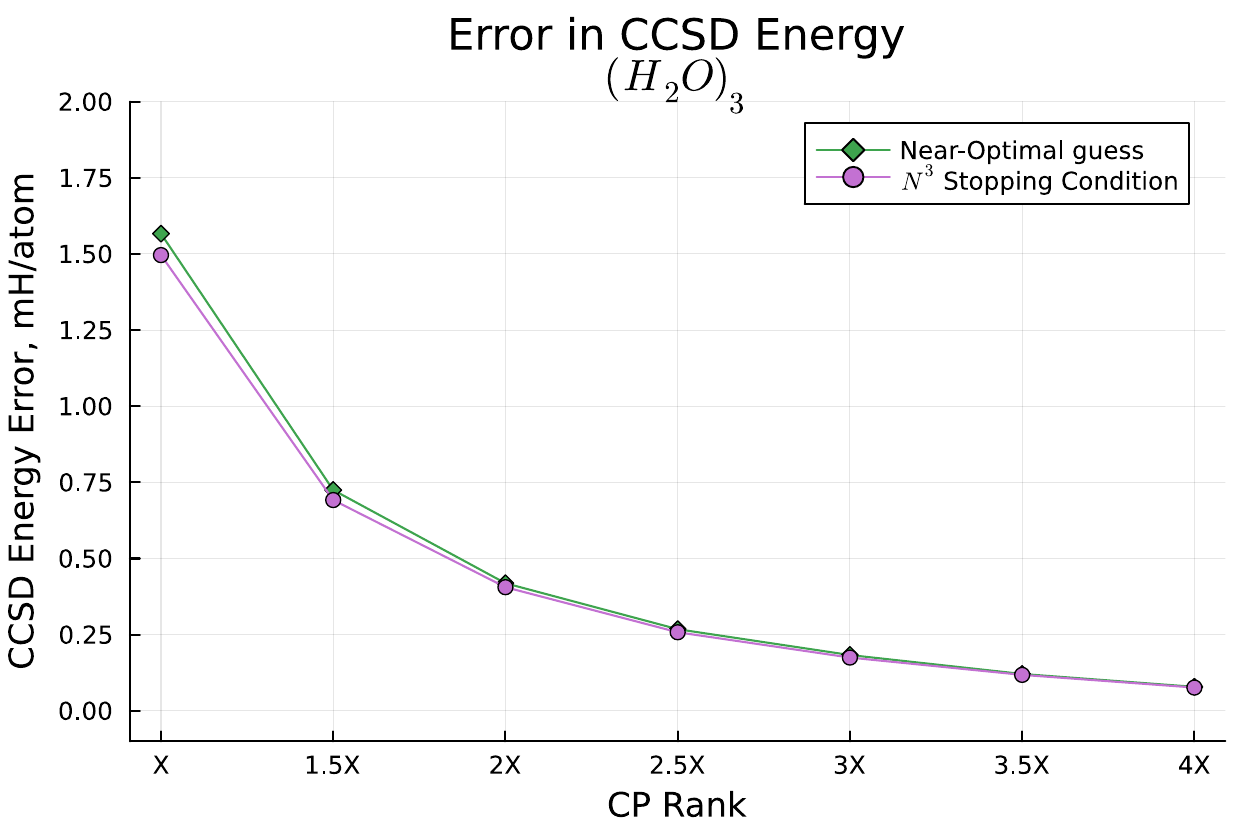}
        \caption{}
        \label{fig:cheap_abs_3w}
    \end{subfigure}\hfill
    \begin{subfigure}{0.49\textwidth}
        \includegraphics[width=\linewidth]{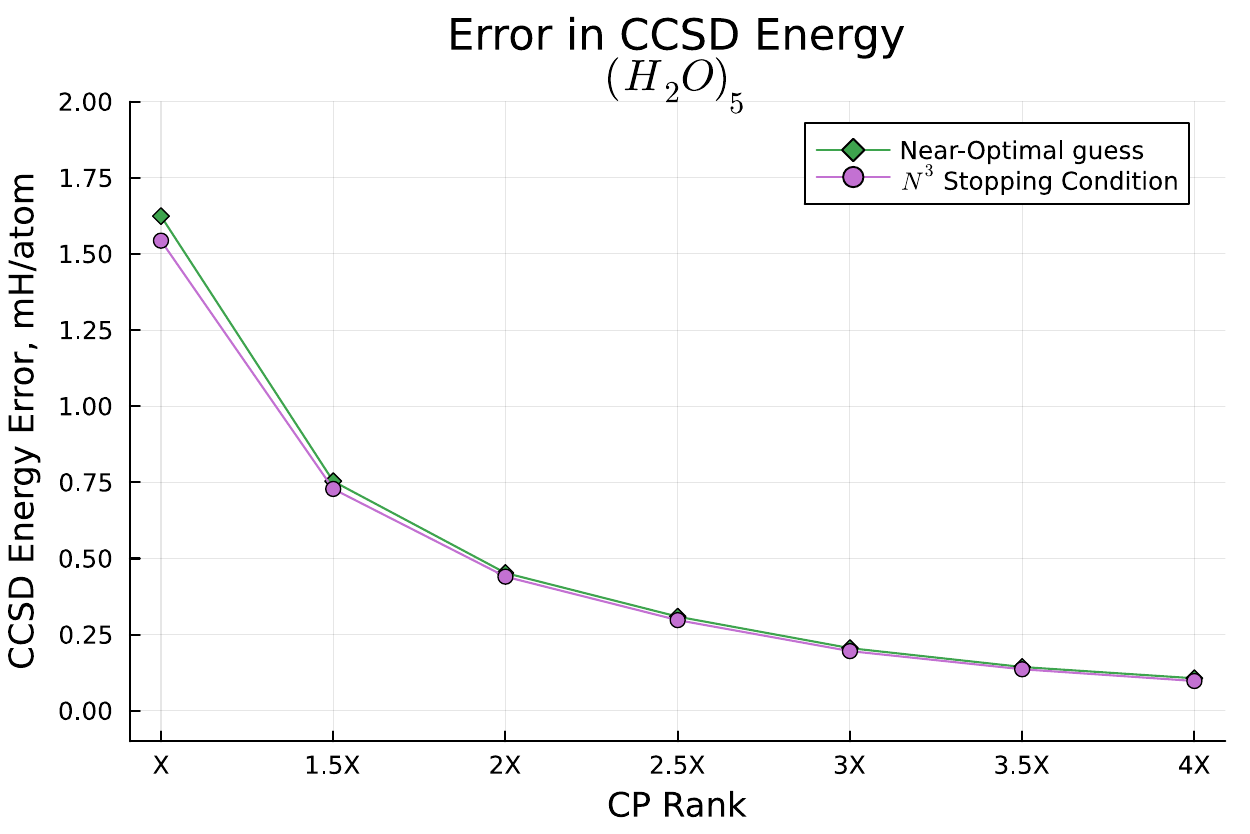}
        \caption{}
        \label{fig:cheap_abs_5w}
    \end{subfigure}\hfill
    \begin{subfigure}{0.5\textwidth}
        \includegraphics[width=\columnwidth]{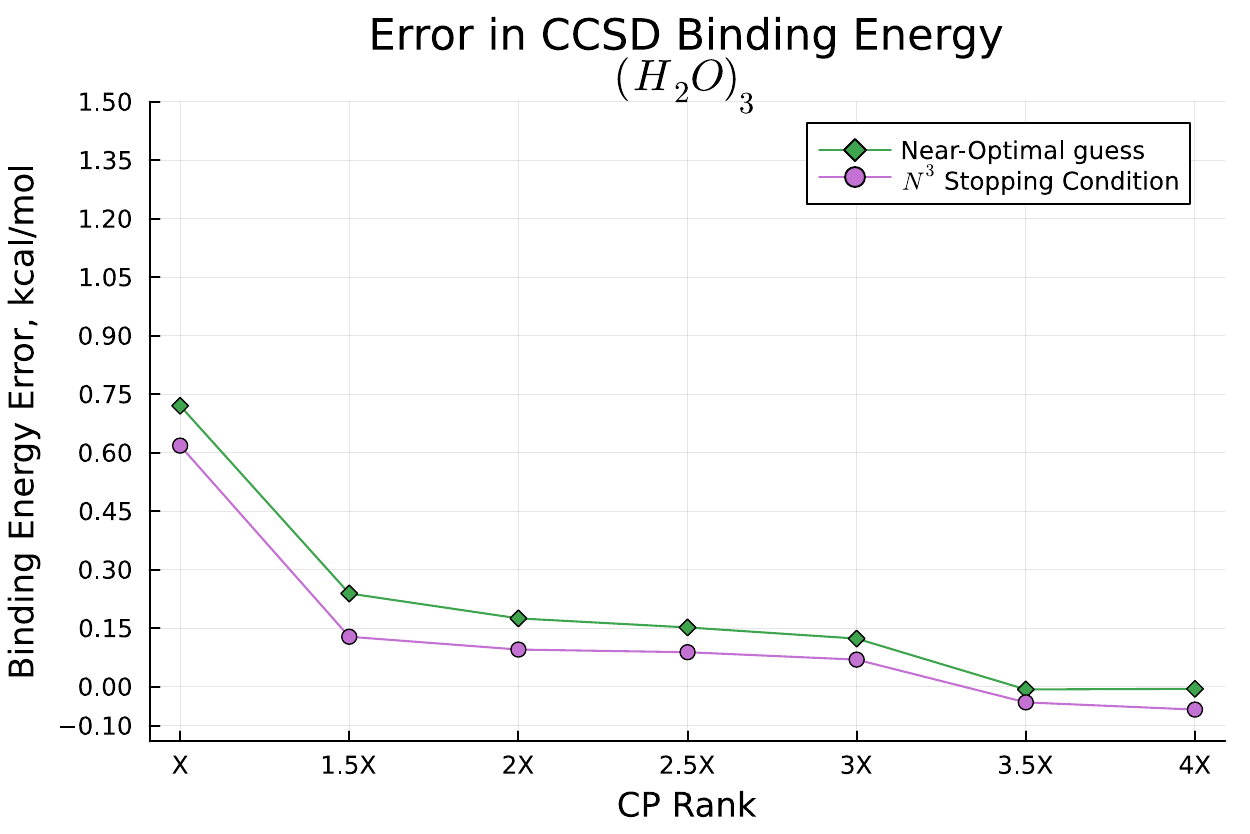}
        \caption{}
        \label{fig:cheap_bind_3w}
    \end{subfigure}\hfill
    \begin{subfigure}{0.49\textwidth}
        \includegraphics[width=\linewidth]{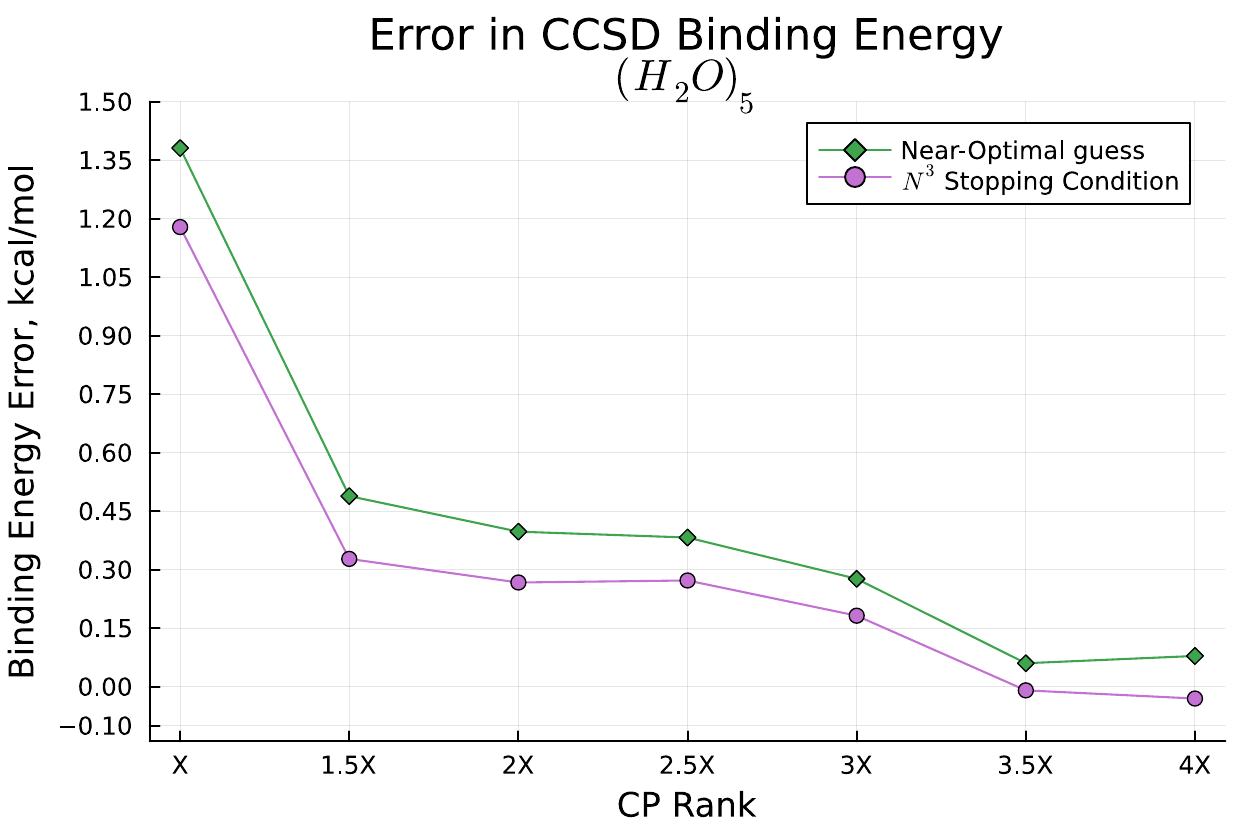}
        \caption{}
        \label{fig:cheap_bind_5w}
    \end{subfigure}\hfill
\caption{Comparing the absolute CCSD energy error for a (a) 3 and (b) 5 water molecule cluster using the canonical and reduced-scaling stopping CP-PPL condition. 
Comparing the binding energy error for a (c) 3 and (d) 5 water molecule cluster using the canonical and reduced-scaling CP-PPL stopping condition.} \label{fig:n3Energy}
\end{figure}
With \cref{fig:cheap_als_conv}, we show the convergence behavior of the CP-PPL using the reduced-scaling stopping condition.
One can see that the behavior of this convergence is nearly identical to the convergence of the "Near-Optimal guess" with the canonical stopping condition, i.e., \cref{fig:no_guess_als_conv}. 
Finally, we analyze the accuracy of the CCSD results using this new stopping condition in \cref{fig:n3Energy}.
This figure illustrates that the reduced-scaling initial guess scheme realizes nearly identical results for absolute and relative energy quantities compared to the canonical stopping condition.
These results demonstrate that our choice of $\epsilon$ is reasonable and that the new convergence criteria can be effectively used by the CP-PPL optimization scheme.
Next we analyze the scaling of the rank of the CP-PPL approximation.

\begin{figure}
    \begin{subfigure}{0.5\textwidth}
        \includegraphics[width=\columnwidth]{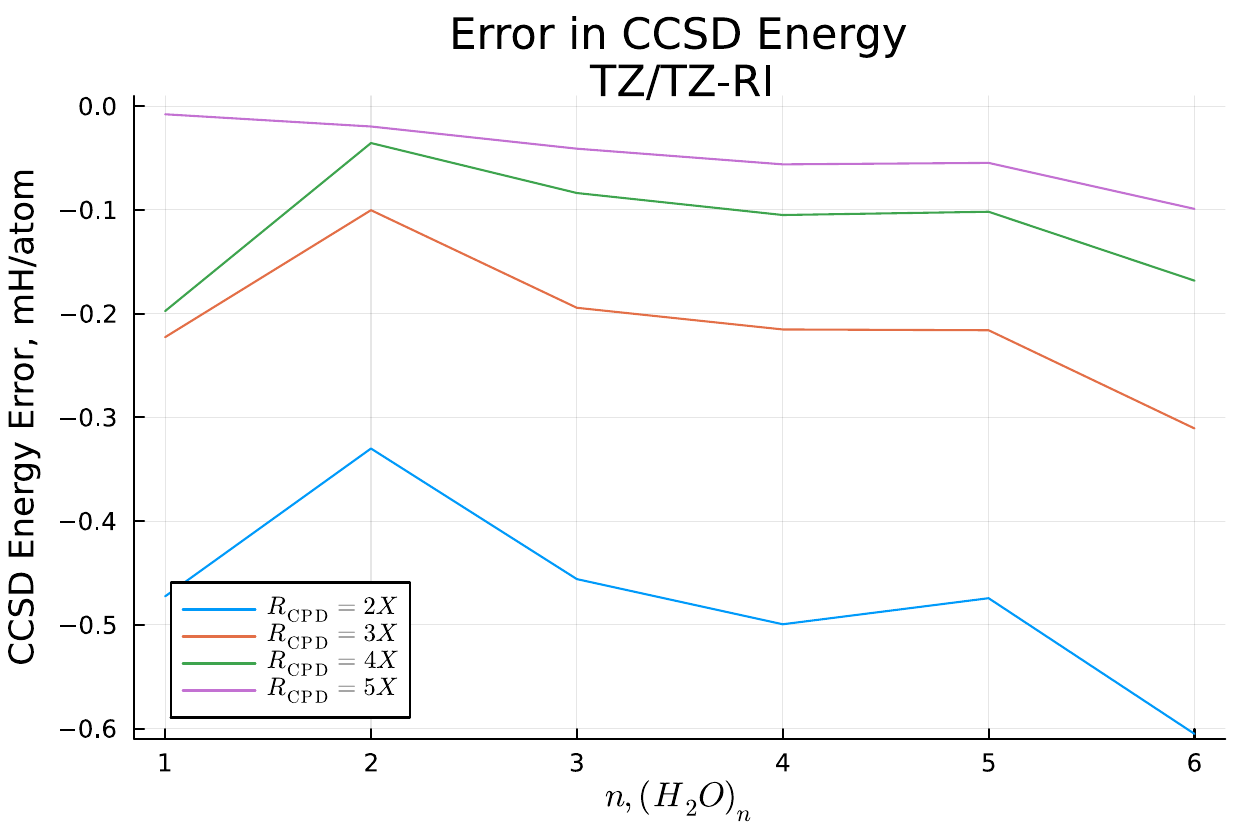}
        \caption{}
        \label{fig:tz-per-atom}
    \end{subfigure}\hfill
    \begin{subfigure}{0.5\textwidth}
        \includegraphics[width=\columnwidth]{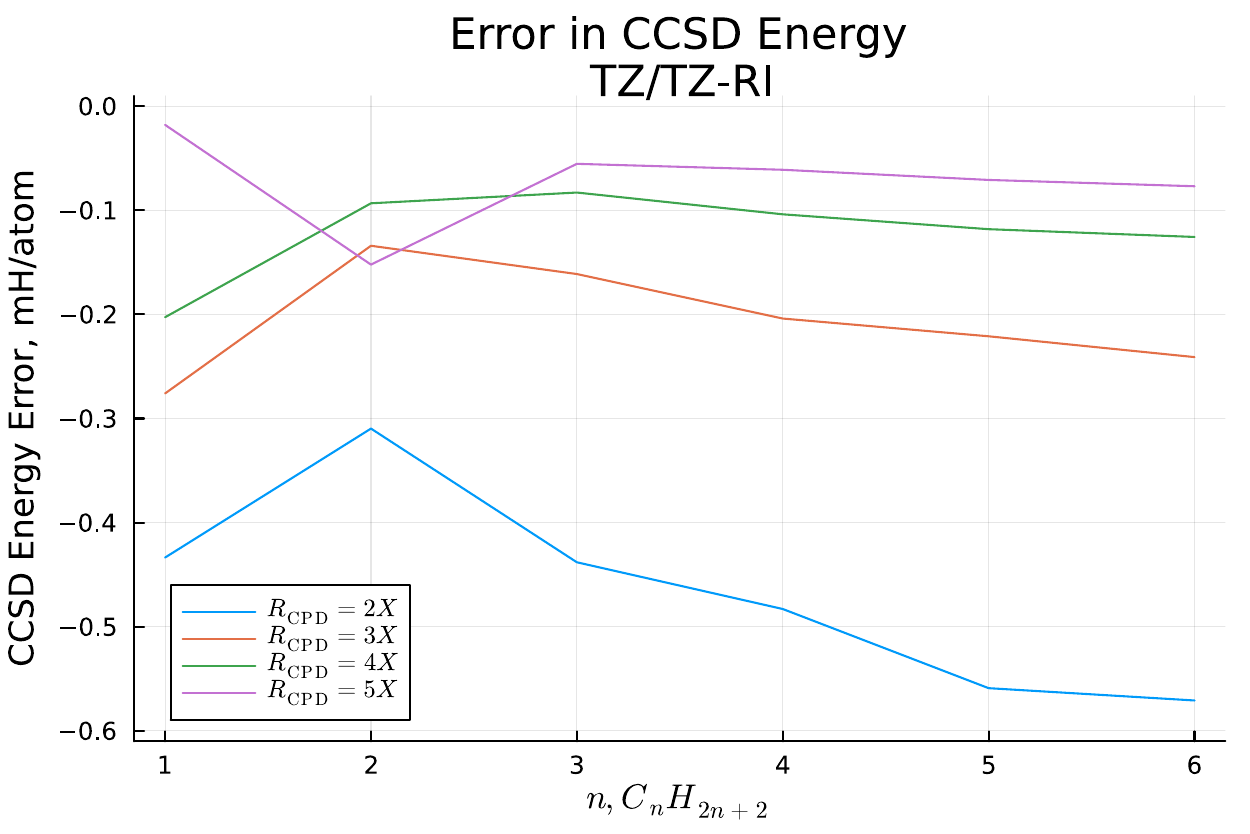}
        \caption{}
        \label{fig:tz-alk-per-atom}
    \end{subfigure}\hfill
\caption{Absolute error in CP-PPL CCSD energy per non-hydrogenic atom for (a) water clusters in the TIP4P geometry using the TZ/TZ-RI basis sets and (b) for alkane chains in the TZ/TZ-RI basis.}
\label{fig:scaling_errors}
\end{figure}
\subsection{CP-PPL Rank Scaling Models}
Here we analyze the accuracy of the CP-PPL approximated CCSD calculation against systematically increasing system size to determine the scaling of the CP rank for this tensor-network.
In \cref{fig:scaling_errors} we plot the absolute error in the CCSD energy using fixed values of CP rank across a set of water clusters and alkane chains in the TZ/TZ-RI basis. 
One can see that, for the most part, increasing the CP rank systematically improves the accuracy of the CP-PPL approximated CCSD method.
One exception to this trend is the \ce{C2H8} molecule with a CP rank of $5X$.
We can occasionally find this kind of inconsistency when decomposing small problems to large CP rank; it is most likely related to the optimization of an over-parameterized alternating least squares problem.

\begin{figure}
    \begin{subfigure}{0.5\textwidth}
        \includegraphics[trim=2cm 3cm 3cm 3cm, clip, width=\columnwidth]{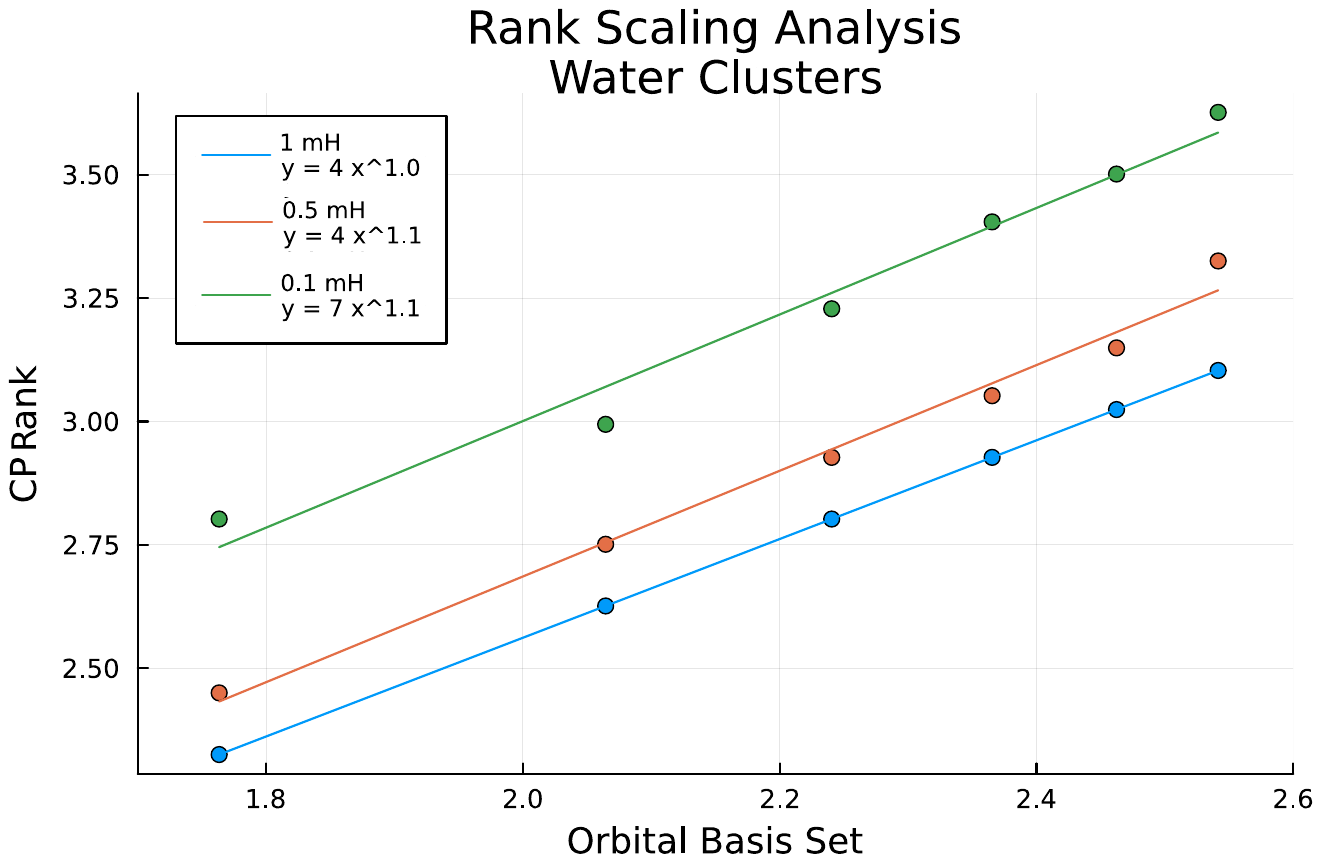}
        \caption{}
        \label{fig:tz-wat-scale}
    \end{subfigure}\hfill
    \begin{subfigure}{0.5\textwidth}
        \includegraphics[trim=2cm 3cm 3cm 3cm, clip,width=\columnwidth]{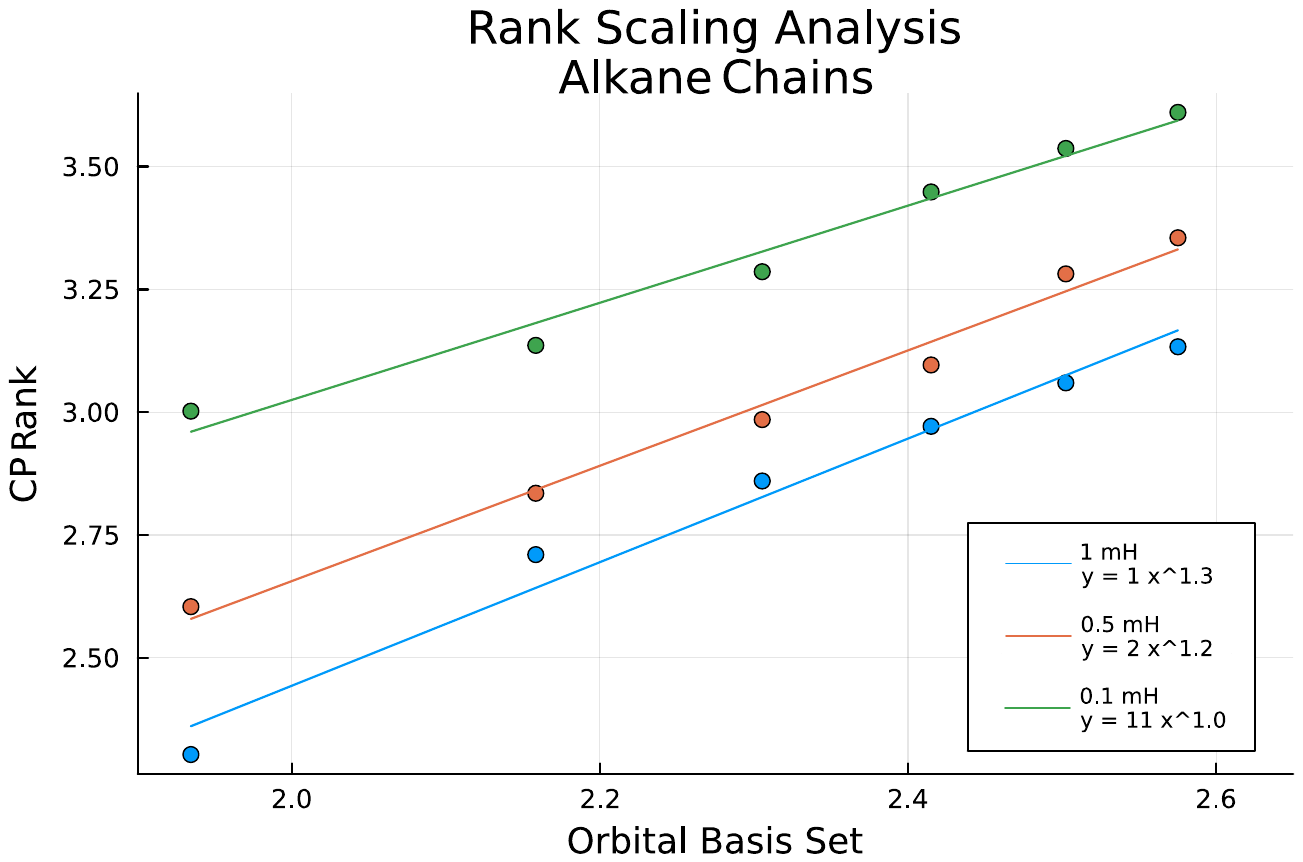}
        \caption{}
        \label{fig:tz-alk-scale}
    \end{subfigure}\hfill
\caption{Modeling the growth of the CP rank with system size for (a) water molecule clusters and (b) alkane chains in the TZ/TZ-RI basis using three different error thresholds}
\label{fig:scaling_model}
\end{figure}

Furthermore, there appears to be a slight increase in CCSD energy error, with increasing system size.
Although, because the tested systems are rather small, it is hard to determine if this trend will continue for larger system sizes. 
Because this work is a preliminary study, we leave that investigation to future works on this matter.
Using the data from \cref{fig:scaling_errors}, we plot the CP rank versus system size for various fixed error thresholds in \cref{fig:scaling_model}.
The models computed from the data in \cref{fig:scaling_model} illustrate that the rank of the  CP-PPL approximation is nearly linear, falling between $N$ and $N^{1.3}$.
This scaling is consistent with other works studying low-rank decompositions of tensors in CCSD such as Benedikt et al.\cite{Benedikt:2013:JCP} and Schutksi et al.\cite{Schutski:2017:JCP}

\subsection{CP-PPL approximated CCSD Performance}
\begin{figure}[t!]
    \includegraphics[width=\columnwidth]{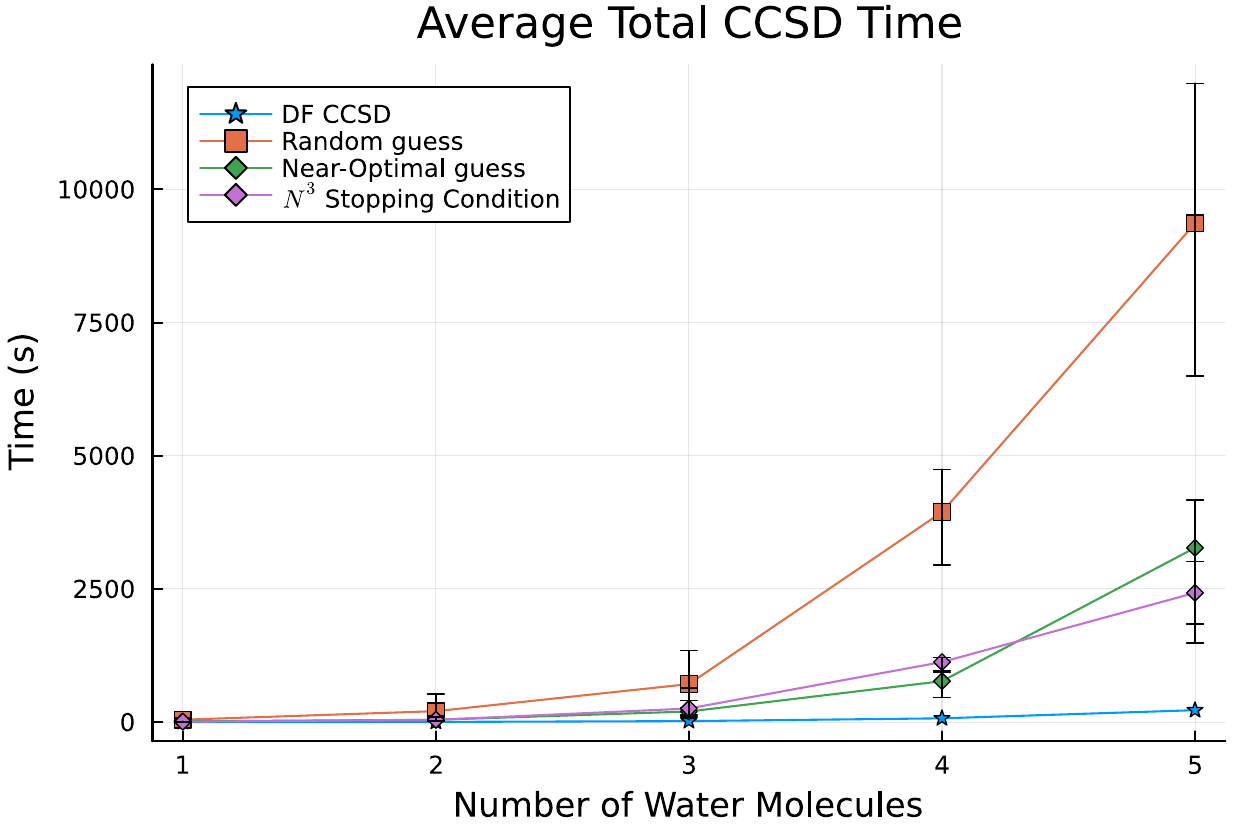}
        \caption{Total time to compute the CCSD energy for water clusters in the DZ-F12/aVDZ-RI basis. CP-PPL approximated CCSD calculations were averaged over all CP rank values and the error bars represent the maximum and minimum recorded times.}
        \label{fig:n3avgtiming}
\end{figure}
\begin{figure}
    \begin{subfigure}{0.5\textwidth}
        \includegraphics[width=\columnwidth]{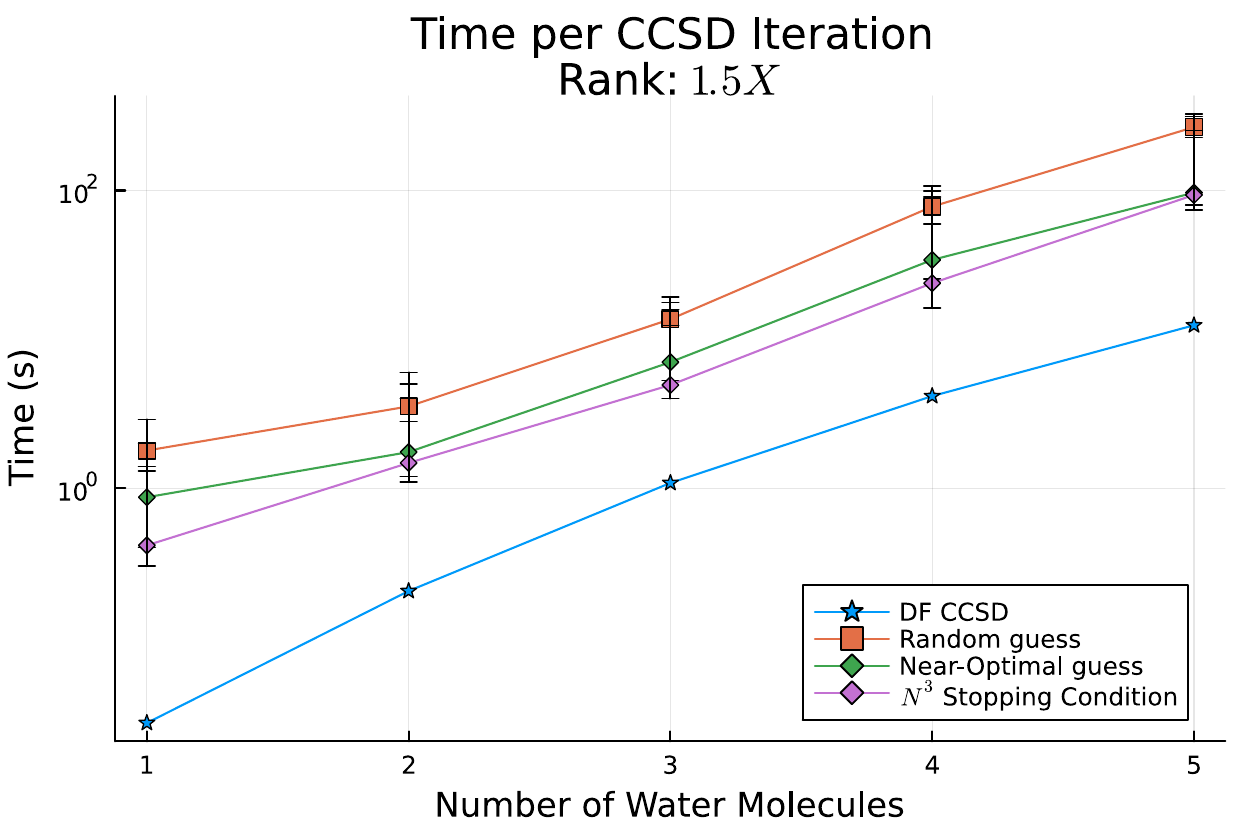}
        \caption{}
        \label{fig:periter1.5}
    \end{subfigure}\hfill
    \begin{subfigure}{0.49\textwidth}
        \includegraphics[width=\linewidth]{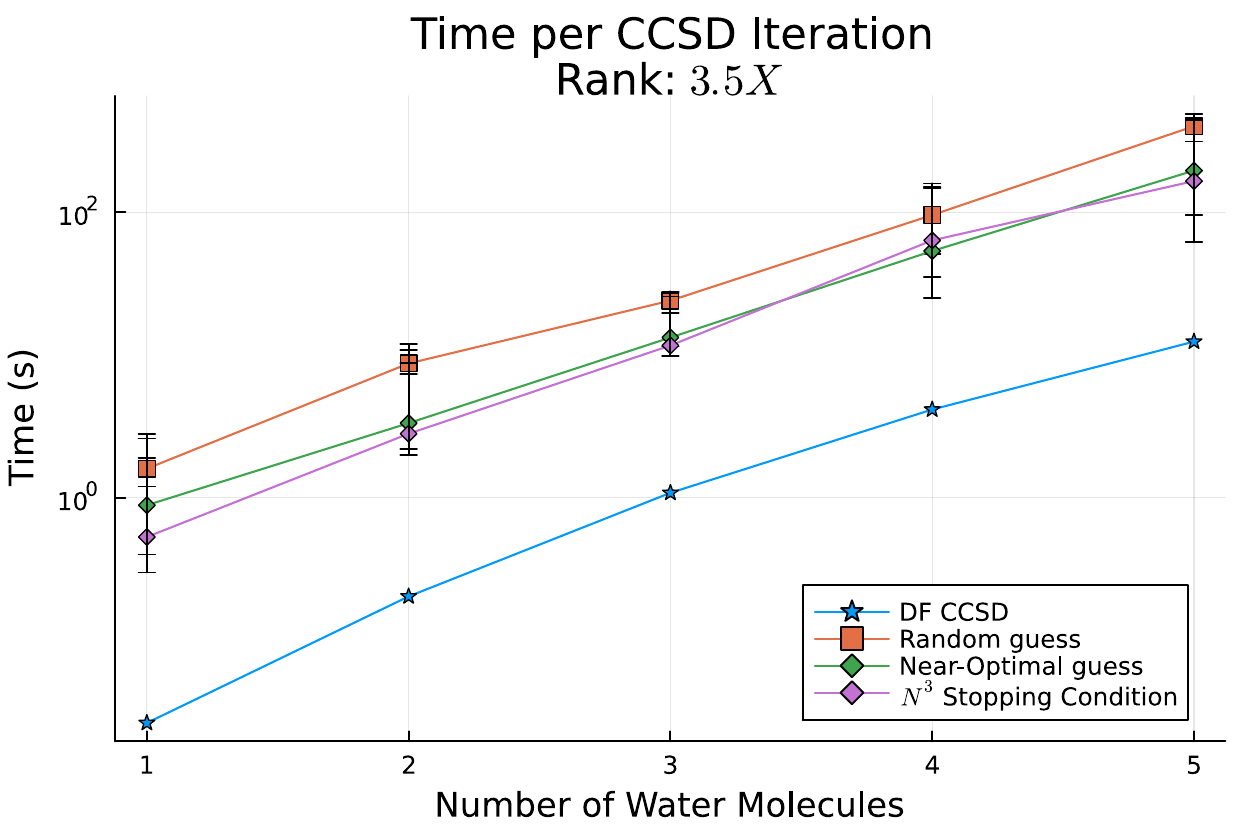}
        \caption{}
        \label{fig:periter3.5}
    \end{subfigure}\hfill
\caption{CCSD per iteration cost versus number of water molecules in the DZ-F12/aVDZ-RI basis. Results were compiled using a CP rank of (a) $1.5X$ and (b) $3.5X$. Error bars represent maximum and minimum per iteration time.} \label{fig:n3periter}
\end{figure}

Finally, because the CP-PPL is technically a reduced scaling approach, we show the wall-time cost of the CP-PPL approximated CCSD compared to a canonical DF CCSD optimization in \cref{fig:n3avgtiming,fig:n3periter}.
As discussed earlier, the point of this work is not to realize a wall-time improvement over the canonical contraction of the PPL diagram although, for completeness, we discuss the theoretical crossover.
This matrix-free CPD optimization has a leading cost of $(V^2O^2 R)$ per CPD ALS update and an additional $(V^2O^2 R)$ cost to reconstruct the PPL tensor after the CPD optimization.
The canonical PPL has a computational scaling of $(V^4O^2)$.
Theoretically there should be a crossover in cost when $V^2 > R(N_{ALS} + 1)$ where $N_{ALS}$ is the number of ALS iterations.
In this work, we have shown that a CP rank of $4X$ is sufficiently accurate and an acceptable value of $X$ is on the order of $2V$.
Therefore, this approach should find advantage over the canonical PPL when $V > 8 (N_{ALS} + 1)$.
The number of ALS iterations depends on the ALS stopping condition and initial guess but is typically between 8 and 30, which would put the theoretical crossover point between $[65, 241]$ unoccupied orbitals.
However, because the CPD has a relatively high prefactor, as discussed in \cref{sec:CPD-PPL}, it is difficult to realize this cross-over. 
In this direction, we seek to improve the timing of the code by distributing the optimization algorithm and utilizing randomized/interpolation based methods to further reduce the prefactor of the CPD of tensor-networks.
Also, we are investigating the application of this decomposition-based tensor-network contraction technique to networks that can be computed more efficiently.

\section{\label{sec:conclusions} Conclusions}
In this study, we demonstrate that it is possible to approximately contract tensor-networks through the optimization of a low-rank tensor decomposition and, by doing so, make the error in the contraction a controllable parameter of the decomposition.
This idea is not limited to a specific tensor-network or decomposition topology.
Here, as a case study, we investigate the approximation of the DF-approximated PPL tensor-network in the CCSD method using the CPD.
We find that computing a CPD approximation of the PPL tensor-network can be done accurately with a relatively small CP rank that grows nearly-linearly with chemical system size.
These results complement our previous results from the investigation into the order-4 CPD of the coulomb integral tensor-network.\cite{VRG:pierce:2023:JCTC,Pierce:2025:JCTC}
Furthermore, we utilize the iterative nature of CCSD to construct an effective CPD initial guess strategy that minimizes the cost of the CPD optimization.
Unfortunately, we find that the approximation of the PPL ladder with this new scheme can have a significant effect on the optimization of the CCSD wavefunction.
To overcome this problem in a standard Jacobi-DIIS CCSD optimization loop, measures must be taken to ensure a consistent CPD approximation.
We believe that these results warrant an investigation into CCSD solvers with improved stability. 

In future works, we will study methods to improve the approximation of tensor-networks computed within iterative optimization schemes like coupled cluster.
For example, we are currently investigating an approach to reduce the computational storage complexity of the CCSD residual tensor by computing its approximation in a low-rank tensor decomposition format.
We also seek to replace more complex and higher-order tensor-networks such as the particle-particle ladder diagram found in the rank-reduced coupled cluster with single, double and triple excitations (CCSDT) method.\cite{VRG:lesiuk:2020:JCTC}
Finally, we look to use more elaborate strategies to better and more quickly decompose tensor-networks. 
These strategies include using sparsity constraints, utilizing low-cost nonlinear optimization schemes, and using random/interpolation methods to more quickly approximate the gradient of tensor-networks.

Supporting Information: xyz files for the alkane chains and additional result figures for CP-PPL convergence and CP rank scaling.

\begin{acknowledgement}
This work was supported by the Flatiron Institute and the Simons Foundation. We also acknowledge the Scientific Computing Core (SCC) at Flatiron Institute\\ (https://www.simonsfoundation.org/flatiron/scientific-computing-core/) for providing computational resources and technical support that have contributed to the results reported within this paper. Furthermore it is important to acknowledge Dr. J Kaye and Dr. A. Dawid at the Flatiron Institute for their valuable discussions and the author thanks Mina Mandic for contributing in the early stages of the project.
\end{acknowledgement}

\bibliography{kmprefs_updated}

\end{document}